\documentclass[showpacs,preprintnumbers,amsmath,amssymb,twocolumn,nofootinbib,nobibnotes,longbibliography]{revtex4-1}
\usepackage{color}
\usepackage[colorlinks,urlcolor=blue]{hyperref}
\usepackage{epstopdf}
\usepackage{graphicx}

\usepackage[english]{babel}
\usepackage{tabularx}

\usepackage{dcolumn}
\usepackage{multirow}
\usepackage{bm}
\usepackage{upgreek}

\begin{document}

\title{
Skew Scattering and Side Jump Drive Exciton Valley Hall Effect\\ in Two-Dimensional Crystals
}

\author{M.~M.~Glazov}
\author{L.~E.~Golub} 	
\affiliation{Ioffe Institute,  	194021 St.~Petersburg, Russia}

\begin{abstract}
Exciton Valley Hall effect is the spatial separation of the valley-tagged excitons by 
a drag force. Usually, the effect is associated with the anomalous velocity acquired by the particles due to the Berry curvature of the Bloch bands. Here we show that the anomalous velocity plays no role in the exciton valley Hall effect, which is governed by the side-jump and skew scattering. 
We develop microscopic theory of the exciton valley Hall effect in the presence of synthetic electric field and phonon drag and calculate all relevant contributions to the valley Hall current also demonstrating the cancellation of the anomalous velocity. The sensitivity of the effect to the origin of the drag force and to the scattering processes is shown. We extend the drift-diffusion model to  account for the valley Hall effect and calculate the exciton density and valley polarization profiles.
\end{abstract}

\maketitle

\emph{Introduction.} Atomically-thin transition metal dichalcogenide monolayers (TMDC MLs) form a basis for van der Waals heterostructures, a novel versatile semiconductor platform with fascinating fundamental physics and wide prospects for applications~\cite{Kolobov2016book,Geim:2013aa}. TMDC MLs demonstrate particularly strong optical response dominated by the robust tightly-bound excitons~\cite{Mak:2010bh,PhysRevLett.113.026803,RevModPhys.90.021001} which demonstrate particularly strong light-matter coupling~\cite{Schneider:2018aa,Krasnok:18}. Transport properties of excitons in two-dimensional (2D) semiconductors attract increasing interest~\cite{Mouri:2014a,Kato2016,doi:10.1021/acs.jpclett.7b00885,Cadiz:2018aa,Wang:2019aa,leon2019hot,zipfel2019exciton} due to unusual linear and nonlinear effects like halo formation~\cite{PhysRevLett.120.207401,Perea-Causin:2019aa,PhysRevB.100.045426} and prospects to observe quantum effects such as weak localization of excitons~\cite{PhysRevLett.124.166802}.

Excitons in 2D TMDC possess valley degree of freedom and the chiral optical selection rules allow one to address the states in individual $\bm K_+$ and $\bm K_-$ valleys by the circularly polarized light and control the superpositions of the states by linearly polarized radiation~\cite{Sallen:2012qf,Jones:2013tg,Yu:2014fk-1,Yu30122014,PSSB:PSSB201552211,PhysRevB.92.075409,PhysRevLett.117.187401}. It opens up a possibility to observe and study the transport of valley-tagged excitons~\cite{Rivera688,Lundt:2019aa,Unuchek:2019aa} and, in particular, the valley Hall effect (VHE)~\cite{PhysRevLett.115.166804,PhysRevB.93.045414,Yu:2014fk-1,Onga:2017aa,Huang:2020aa}, see also~\cite{PhysRevB.78.205201}.

\begin{figure}[h]
\includegraphics[width=0.8\linewidth]{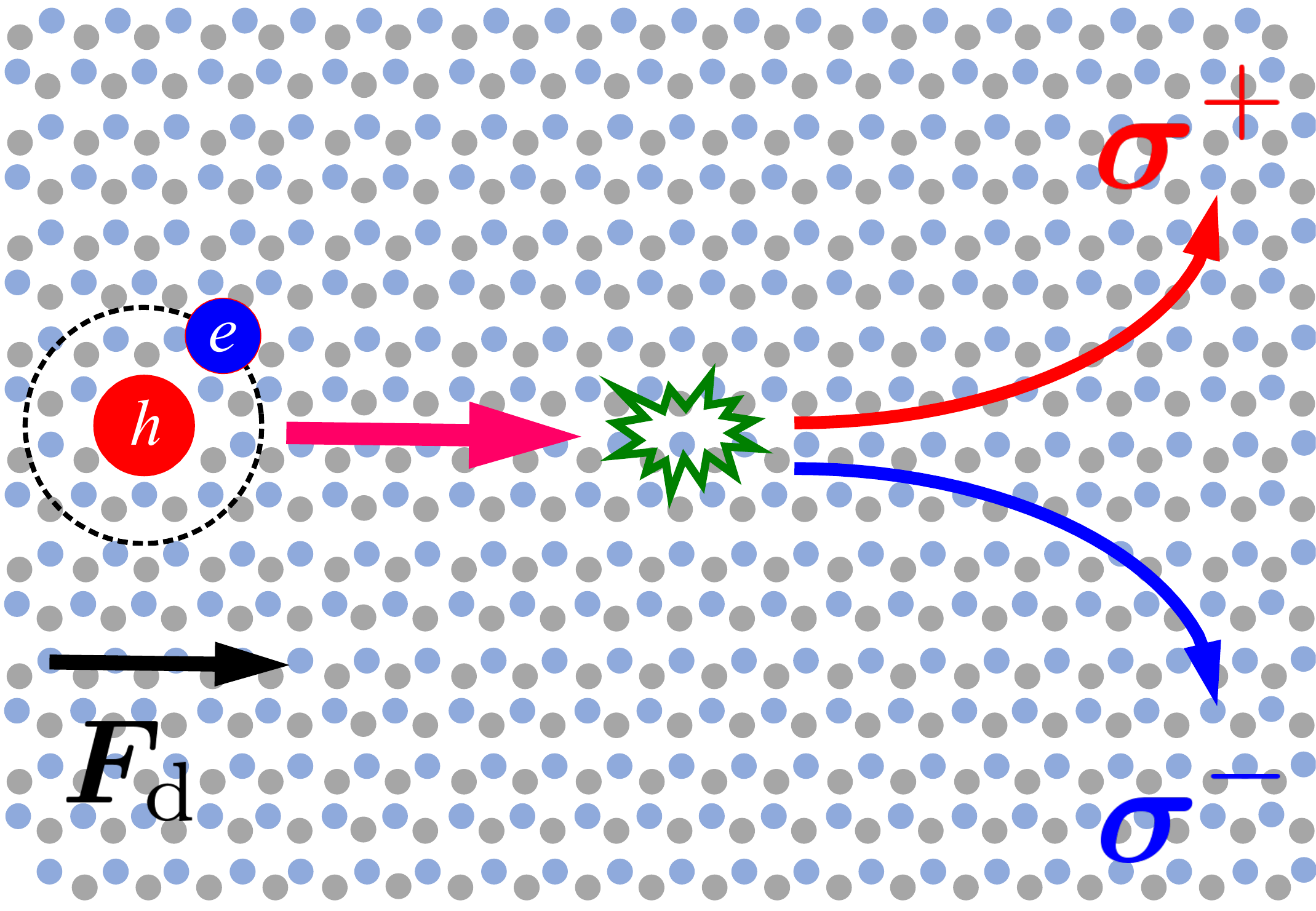}
\caption{Schematics of the 
VHE for excitons, $\bm F_d$  shows the drag force and arrows denote the propagation direction of the excitons in $\bm K_\pm$ valleys. This figure illustrates the skew scattering where the excitons in $\bm K_\pm$ are separated due to asymmetric scattering by impurities or phonons.}\label{fig:VHE}
\end{figure}

The VHE results in the generation of opposite fluxes of $\bm K_+$ and $\bm K_-$ excitons in the presence of drag force $\bm F_d$ which produces a net unidirectional flow of the particles, Fig.~\ref{fig:VHE}. The exciton valley flux $\bm i_{v}=(\bm i_+ - \bm i_-)/2$ with $\bm i_\pm$ being the fluxes in the $\bm K_\pm$ valleys reads~\cite{note:static:linear}
\begin{equation}
\label{iVHE}
\bm i_v = {\chi N} [\hat{\bm z} \times \bm F_d],
\end{equation}
with the constant $\chi$ and $N$ being the total exciton density. 
The chiral structure of states in each valley can be viewed as an effective magnetic field which winds excitons similarly to the cyclotron winding of electrons in a real magnetic field. Due to the time-reversal symmetry of the TMDC MLs~\cite{Xiao:2012cr,2053-1583-2-2-022001} this effective magnetic field is opposite in $\bm K_+$ and $\bm K_-$ valley, resulting in the opposite winding directions and giving rise to the  valley Hall (VH) flux, Eq.~\eqref{iVHE} and spatial separation of excitons, Fig.~\ref{fig:VHE}. {Equation~\eqref{iVHE} plays a role of material relation for exciton transport. }The microscopic mechanisms of the exciton VHE are insufficiently studied and, similarly to the electron VHE, are usually related to the anomalous velocity induced by the force $\bm F_d$~\cite{Xiao:2012cr,PhysRevLett.115.166804,Yu:2014fk-1,Onga:2017aa,PhysRevLett.122.256801,Gianfrate:2020aa,Huang:2020aa}.

Here we uncover that exciton VHE in 2D semiconductors is caused by the skew-scattering and side-jump mechanisms, resulting from, respectively, asymmetric exciton scattering by phonons or impurities and exciton displacement in the course of scattering. The anomalous velocity contribution is compensated or absent depending on the origin of the drag force $\bm F_d$. We present the microscopic theory of the exciton VHE  and calculate $\chi$ in Eq.~\eqref{iVHE} for relevant mechanisms of exciton drag and scattering. We demonstrate that the resulting VH current 
depends both on the origin of the drag force and on the details of the scattering. We also extend the drift-diffusion description of excitonic transport~\cite{PhysRevLett.120.207401,PhysRevLett.120.207401,PhysRevB.100.045426,zipfel2019exciton} to account for the VHE.

\emph{Model.}  The Coulomb-bound electron-hole pair excited by the $\sigma^\pm$-polarized photon is described by the Bloch function~\cite{ivchenko05a,PhysRevLett.101.106401,PhysRevB.95.035311}
\begin{equation}
\label{exciton:bloch}
\Psi_{exc,\pm} = e^{\mathrm i \bm K \bm R} \sum_{\bm k} e^{\mathrm i \bm k \bm \rho} F(\bm k) |e,\bm k_e,\pm \rangle|h,\bm k_h, \mp \rangle.
\end{equation}
 Here $\bm R$ and $\bm K$ is the exciton center of mass coordinate and wavevector, $\bm \rho$ and $\bm k$ is the electron-hole relative motion coordinate and wavevector, $F(\bm k)$ is the Fourier transform of the relative motion envelope,
and $|e,\bm k_e,\pm \rangle$ ($|h,\bm k_h, \mp \rangle$) are the electron (hole) Bloch amplitudes and the corresponding wavevectors are given by $\bm k_e = (m_e/M)\bm K + \bm k$, $\bm k_h = (m_h/M) \bm K - \bm k$ with $m_e$, $m_h$ and $M=m_e+m_h$ being the electron, hole and exciton mass, respectively. Equation~\eqref{exciton:bloch} is valid at the exciton binding energy $E_B$ and its kinetic energy $\mathcal E_K = \hbar^2 K^2/2M$ being much smaller than the band gap $E_g$ and the distance to other bands {see SI~\cite{note:suppl} for derivation and discussion of the model applicability and its extensions to the multiband case.} 
{Equation~\eqref{exciton:bloch} allows} us to calculate the excitonic Berry curvature in the $\bm K_+$ valley~\cite{PhysRevLett.101.106401,PhysRevB.100.121405,note:suppl}
\begin{equation}
\label{curvature}
\bm{\mathcal F} = -2  \left[\xi_e\left(\frac{m_e}{M}\right)^2  + \xi_h \left(\frac{m_h}{M}\right)^2\right]\hat{\bm z},
\end{equation}
with $\hat{\bm z}$ being the unit vector along the ML normal, $\xi_e$ and $\xi_h$ being corresponding parameters describing the electron and hole Berry curvature, 
intraband contribution to the exciton position operator
\begin{equation}
\label{position}
\bm \Omega_{\bm K} = \frac{1}{2} [\bm{\mathcal F} \times \bm K],
\end{equation}
and the matrix element of the exciton scattering [cf.~\cite{2020arXiv200405091G}]
\begin{multline}
\label{scatt}
M_{\bm K'\bm K} = V_c(\bm Q) - V_v(\bm Q) \\
{- \mathrm i \xi \left[V_c(\bm Q)\frac{m_h}{M} - V_v(\bm Q) \frac{m_e}{M} \right][\bm K'\times \bm K]_z.}
\end{multline}
Here $\bm Q= \bm K' - \bm K$ is the scattering wavevector, $V_c(\bm Q)$ and $V_v(\bm Q)$ are the Fourier components of the scattering potential acting in the conduction and valence bands, respectively; the hole potential $V_h = - V_v$; Eq.~\eqref{scatt} is written disregarding remote bands contribution to the Berry curvature, $\xi = \xi_e =\xi_h$, {and we have omitted valley-independent ${\propto}(\bm K\cdot \bm K')$ correction to Eq.~\eqref{scatt}~\cite{note:suppl}}. In the $\bm K_-$ valley the signs of $\xi_e$, $\xi_h$ and $\xi$ in Eqs.~\eqref{curvature} and \eqref{scatt} 
are inverted. 

The unidirectional flux of the excitons can be provided by several sources, including (i) synthetic static field acting on the exciton as a whole~\cite{PhysRevLett.101.106401,PhysRevB.100.121405}, (ii) phonon drag or wind~\cite{keldysh_wind,sibeldin_bagaev_wind,PhysRevB.100.045426}, (iii) exciton  density or  temperature gradient~\cite{Onga:2017aa,Perea-Causin:2019aa}. Here we consider in detail two first options and analyze the exciton 
VHE caused by static field resulting, e.g., from an inhomogeneous strain in the sample, and by the phonon drag{, the option (iii) is addressed in SI~\cite{note:suppl} and shown to be essentially similar to the option (i)}. {We calculate the exciton valley current in the linear in $\bm F_d$ approximation. In the cases (i) and (ii)} 
 the microscopic origin of the force $\bm F_d$ acting on the excitons, Fig.~\ref{fig:VHE} is qualitatively different. Indeed, the inhomogeneous strain produces the coordinate-dependent energy shifts of the conduction and valence bands which results in the variation of the exciton potential energy~\cite{note:full:strain},
\begin{equation}
\label{U}
U(\bm R) = (\Xi_c - \Xi_v)[\epsilon_{xx}(\bm R) + \epsilon_{yy}(\bm R)],
\end{equation}
with $\epsilon_{\alpha\beta}$ being the Cartesian components of the strain tensor and $\Xi_c$, $\Xi_v$ being the conduction and valence band deformation potentials. For the constant strain gradient the synthetic force reads
\begin{equation}
\label{Fd:U}
\bm F_d^{(s)} = - \bm \nabla U(\bm R).
\end{equation}
We stress that this is a potential force because it corresponds to the gradient of the exciton potential energy, Eq.~\eqref{U}. The situation is qualitatively different if the excitons are dragged by the non-equilibrium phonons. The net force due to the phonon drag results from the momentum transfer between the phonons and excitons in the collision, while the exciton motion between the collisions is unaffected by phonons. In this case the drag force~\cite{PhysRevB.100.045426,2020arXiv200405091G}
\begin{equation}
\label{Fd:phon}
\bm F_d^{(ph)} = -\frac{\tau_p^{ph}}{{\varrho} \hbar} \left(\frac{M}{\hbar}\right)^2 (\Xi_c - \Xi_v)^2 k_B \bm \nabla T_{\rm latt},
\end{equation}
is not associated with a gradient of any potential energy. In Eq.~\eqref{Fd:phon} $\tau_p^{ph}$ is the phonon momentum relaxation time, ${\varrho}$ is the mass density of the material and $T_{\rm latt}$ is the lattice temperature.
The difference of the origin of the forces in Eqs.~\eqref{Fd:U} and \eqref{Fd:phon} results, as we show below, in the difference of the VHE mechanisms and in the difference in the constant $\chi$ in Eq.~\eqref{iVHE}.

\begin{figure}[h]
\includegraphics[width=0.99\linewidth]{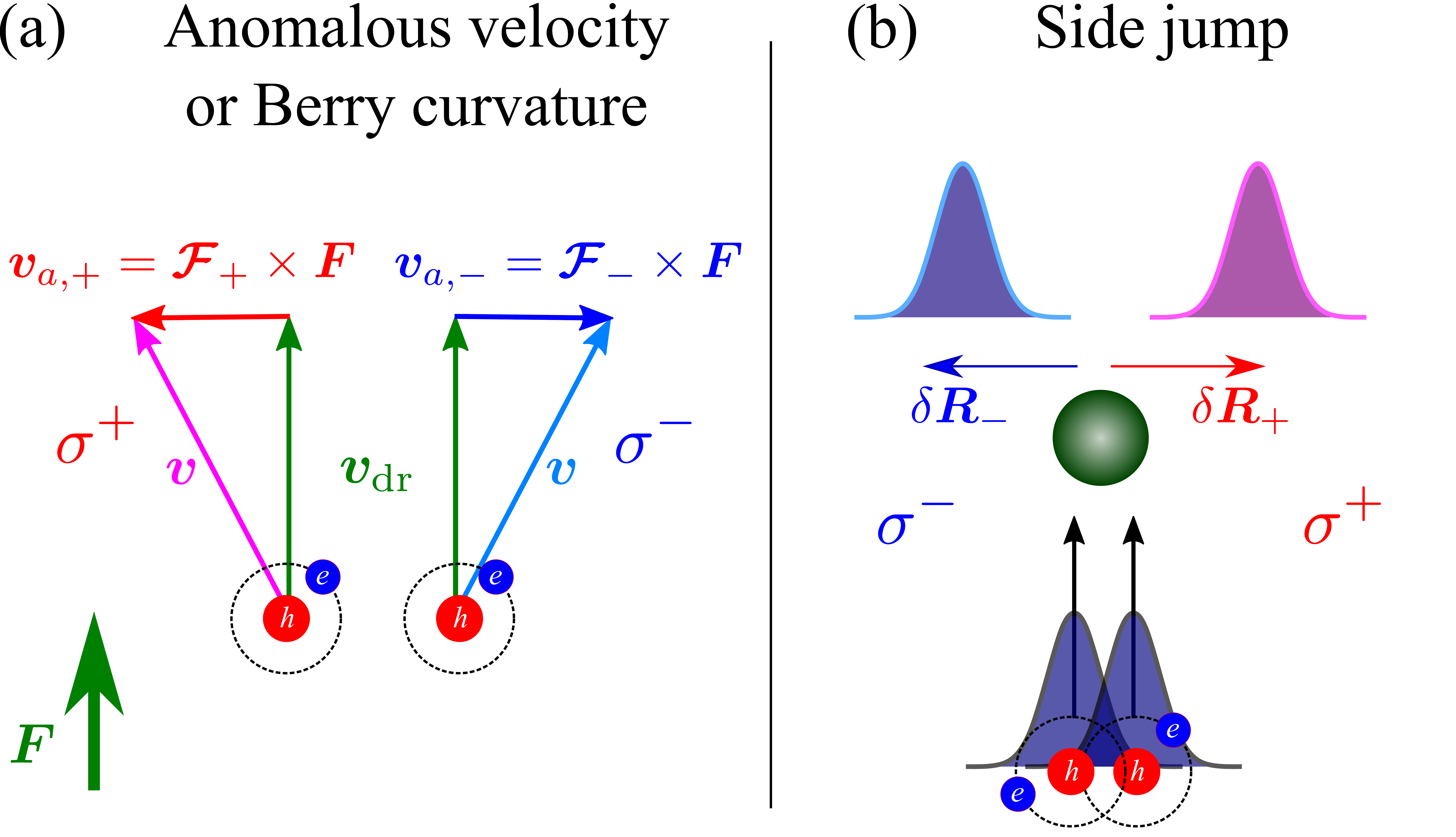}
\caption{Illustration of the anomalous velocity contribution to the VHE (a) and the side-jump contribution (b). We demonstrate the anomalous velocities $\bm v_{a,\pm}$ for excitons in $\bm K_\pm$ valleys (denoted according to the polarization of emission $\sigma^\pm$) and exciton displacements $\delta \bm R_\pm$. The separation of $\sigma^+$ and $\sigma^-$ polarized excitons due to the anomalous velocity and side-jumps are seen.}\label{fig:anom}
\end{figure}

\emph{Mechanisms of VHE}. 
We start with the anomalous contributions illustrated in Fig.~\ref{fig:anom}. The anomalous velocity contribution Fig.~\ref{fig:anom}(a) is readily calculated {from }
Eq.~\eqref{curvature} as~\cite{PhysRevLett.101.106401,PhysRevB.100.121405}
\begin{equation}
\label{j:vhe:av}
\bm i_v^{(av)} = \frac{N}{2} [\bm{\mathcal F}\times \bm F_d^{(s)}].
\end{equation}
Here $\bm F_d^{(s)}$ is the potential force related to the synthetic electric field acting on excitons, Eq.~\eqref{Fd:U}. Interestingly, the scattering rates do not appear in Eq.~\eqref{j:vhe:av}, thus it is commonly assumed that this contribution is the universal and, thus, dominant one for the electronic and excitonic anomalous or VH
effects~\cite{PhysRevLett.101.106401,Onga:2017aa,PhysRevLett.122.256801,Gianfrate:2020aa,PhysRevB.100.121405}. However, this is not true and {similarly to} free charge carriers this contribution is compensated by the so-called side-jump contribution, Fig.~\ref{fig:anom}(b) related to the shifts of excitonic wavepackets under scattering~\cite{dyakonov_book,Sinitsyn_2007,2020arXiv200405091G}. The physical origin of the compensation is clear if one takes into account the anomalous velocity contribution~\cite{note1,nozieresAHE,dyakonov_book} at the exciton scattering by an impurity or phonon. 
The net force acting on the exciton in the steady-state is zero: The force due to external fields or drag is compensated by the friction force due to the impurities and phonons, and total anomalous velocity vanishes.

Microscopically, we calculate the exciton displacement at the scattering as ~\cite{PhysRev.95.1154,belinicher82,Sturman2019,2020arXiv200405091G}
\begin{multline}
\label{X:shifts}
{\bm R_{\bm K'\bm K}}
=  - \xi \frac{V_c(\bm Q)\frac{m_h}{M} - V_v(\bm Q) \frac{m_e}{M} }{V_c(\bm Q) - V_v(\bm Q) } [ (\bm K'-\bm K) \times \hat{\bm z}] \\
+ \frac{1}{2}[\bm{\mathcal F} \times (\bm K'-\bm K)].
\end{multline}
There are two contributions due to the side-jump VH flux of excitons in the case of synthetic field.
One contribution results from the 
shifts, i.e., side-jump accumulation
\begin{equation}
\label{i:vhe:sja}
\bm i_{v}^{(sj,a)} = \sum_{\bm K\bm K'} \bm R_{\bm K'\bm K} {{2\pi\over \hbar}\delta(\mathcal E_K-\mathcal E_{K'})} |M_{\bm K' \bm K}|^2 \delta f_{\bm K},
\end{equation} 
with $\delta f_{\bm K} = -\tau_p (\bm v_{\bm K} \cdot \bm F_d^{(s)}) f_0'(\mathcal E_K)$ being the field-induced anisotropic part of the exciton distribution function responsible for the direct flow of excitons. Here $f_0(\mathcal E_{{K}})$ is the exciton equilibrium distribution function, prime denotes the derivative over energy, $\bm v_{\bm K} = \hbar \bm K/M$ is the exciton velocity,  and $\tau_p$ is the exciton momentum relaxation time. The second one results from the work of the synthetic field at the exciton side-jump and can be reduced to the same form as Eq.~\eqref{i:vhe:sja}: $\bm i_{v}^{(sj,b)} = \bm i_{v}^{(sj,a)}$~\cite{2020arXiv200405091G,note:suppl}.
The total side-jump current acquires the form
\begin{equation}
\label{i:vhe:sj}
\bm i_{v}^{(sj)} = -\frac{N}{2} [\bm{\mathcal F}\times \bm F_d^{(s)}] -  {\frac{\xi}{\hbar}\mathcal A} N [\hat{\bm z} \times \bm F_d^{(s)}],
\end{equation}
{where} we assumed that the scattering is caused by the short-range impurities or acoustic phonons with $V_c$ and $V_v$  independent of the transferred wavevector, ${\mathcal A = {(\nu{m_h} - {m_e})}/[M({\nu - 1})]}$, $\nu = \nu_{\rm imp} \equiv V_c/V_v$ (impurities) or $\nu = \nu_{\rm ph} \equiv\Xi_c/\Xi_v$ (phonons). Combining Eqs.~\eqref{j:vhe:av} and \eqref{i:vhe:sj} we observe that the anomalous velocity contribution vanishes, the resulting current depends on the details of the scattering processes and is given by Eq.~\eqref{iVHE} where
\begin{equation}
\label{i:vhe:anom}
\chi^{(anom, s)}= - \frac{\xi}{\hbar} {\mathcal A.}
\end{equation}
This compensation does not rely on particular form of the Berry curvature~\eqref{curvature}, see Refs.~\cite{2020arXiv200405091G,note:suppl} for detail.

The situation is somewhat different for the anomalous contribution to the current caused by the phonon drag. In this case the only source of the anomalous contribution is the side-jump mechanism, Fig.~\ref{fig:anom}(b) because there is no force acting on excitons between the collisions with phonons. The side-jump current has two contributions, the first one, $\bm i_v^{(sj,rel)}$ arises due to the relaxation of the anisotropic part of excitonic distribution function and can be recast in the form of Eq.~\eqref{i:vhe:sja}. The second contribution arises due to the exciton scattering by anisotropic distribution of phonons~\cite{PhysRevB.100.045426,2020arXiv200405091G} and can be written as
\begin{equation}
\label{i:vhe:sj:aniso}
\bm i_{v}^{(sj,anis)} = \sum_{\bm K\bm K'} \bm R_{\bm K'\bm K} W^{anis}_{\bm K',\bm K} f_0(\mathcal E_k),
\end{equation}
where  $W^{anis}_{\bm K',\bm K}\propto \bm \nabla T_{\rm latt}$ is the rate of the exciton-phonon scattering resulting in the phonon drag. The calculation shows that if the exciton scattering is due to the acoustic phonons only, the sum $\bm i_v^{(sj,rel)}  + \bm i_{v}^{(sj,anis)}\equiv 0$. The origin of this compensation is clear and related, as before, with the steady-state regime of the exciton transport: The excitonic wavepacket displacements due to collisions with phonons resulting in the phonon drag are compensated by the displacements at the collisions resulting in the momentum relaxation. The net VH current arises provided that in addition to the acoustic phonon scattering, the momentum relaxation takes place at static impurities (or other phonons)~\cite{note:suppl}:
\begin{equation}
\label{i:vhe:anom:drag}
\chi^{(anom,d)} = -{\frac{\xi}{{2}\hbar}} (\mathcal A_{\rm imp} - \mathcal A_{\rm ph}) \frac{\tau_p}{\tau_{\rm imp}} 
,
\end{equation}
where $\tau_{\rm imp}$ is the momentum scattering time at the exciton-impurities collisions.

Equations~\eqref{i:vhe:anom} and \eqref{i:vhe:anom:drag} demonstrate that the anomalous contribution to the exciton VH current strongly depends both on the origin of the drag force and on the details of the scattering processes.

Now we calculate the skew-scattering contribution to the exciton VHE, i.e. VH current due to the asymmetric (and opposite in $\bm K_\pm$ valleys) scattering of excitons by impurities or phonons, Fig.~\ref{fig:VHE}. 
Physical origin of the effect is the spin-orbit interaction which produces an effective magnetic field at exciton scattering. 
This 
field has opposite signs in the $\bm K_\pm$ valleys resulting in the VHE.
Quantitatively, the effect is related to the asymmetric $\propto [\bm K'\times \bm K]_{z}$ contribution in the scattering matrix elements~\eqref{scatt}. 
To calculate the skew-scattering contribution to the VHE one has to go beyond the Born approximation in evaluation of the scattering rate: The squared modulus of the matrix element $|M_{\bm K'\bm K}|^2$ does not contain an asymmetric part. In the $\xi$-linear regime the asymmetric contributions to the scattering rate are of interference type and appear (i)~in the third order in $V_c$, $V_v$ and (ii)~in the fourth order as a result of coherent two-impurity or two-phonon scattering~\cite{dyakonov_book,gy61,abakumov72,bs78,2020arXiv200405091G}.

Here we focus on the case where the exciton scattering is dominated by the exciton-acoustic phonon interaction~\cite{note2}. The phonon-induced potential is Gaussian and the third-order contributions $\propto \langle M_{\bm K'\bm K_1} M_{\bm K_1\bm K_2} M_{\bm K_2\bm K}\rangle$    
vanish. Thus, we take into account the two-phonon processes described in detail in the SI~\cite{note:suppl}. 

In the presence of the synthetic field~\eqref{Fd:U}, the interference of the single- and two-phonon processes results in the VH current in the form of Eq.~\eqref{iVHE} with~\cite{2020arXiv200405091G,note:suppl}:
\begin{equation}
\label{i:vhe:skew:2ph}
\chi^{(skew,2ph,s)} = {\frac{\xi}{\hbar}}{\mathcal B} \frac{ k_B T\tilde \Xi_v}{(\Xi_c - \Xi_v)^2 } ,
\end{equation}
where
\[
{\mathcal B =  2\mathcal A_{\rm ph} \left[\frac{m_e}{M} - \frac{m_h}{M}\tilde \nu_{\rm ph} + 2\mathcal A_{\rm ph}(1-\tilde \nu_{\rm ph})\right],}
\]
$\tilde \nu_{\rm ph} = \tilde \Xi_c/\tilde \Xi_v$, and $\tilde \Xi_c$, $\tilde \Xi_v$ are the two-phonon deformation potentials.
We also take into account the coherent two-phonon processes where the intermediate states lie in the same band~\cite{Ado_2015,PhysRevB.96.235148,2020arXiv200405091G} with the result~\cite{note:suppl}
\begin{equation}
\label{i:vhe:skew:2ph:coh}
{\chi^{(skew,coh,s)} =  \frac{\xi}{\hbar}} \mathcal A_{\rm ph}.
\end{equation}

Under the phonon drag conditions we need to take into account additional contribution to the VH current arising from the skew scattering on the anisotropic phonon distribution. Calculation~\cite{note:suppl} shows that:
\begin{equation}
\label{i:vhe:skew:drag}
\chi^{(skew,d)} = \frac{\chi^{(skew,2ph,s)}}{2}+ \frac{\chi^{(skew,coh,s)}}{4}.
\end{equation}

\emph{Results.} The microscopic theory demonstrates that the excitonic VHE  in two-dimensional semiconductors is driven by the skew scattering and side jump effects. 
The side-jump contribution is also crucial for anomalous transport of excitons and should be taken into account along with the skew scattering for the anomalous exciton Hall effect~\cite{2020arXiv200608717K}. 
Importantly, the coherent two-phonon skew scattering and the side jump contributions have similar order of magnitude and in the case of synthetic field $\bm F_d^{(s)}$ exactly compensate each other. Thus, in a synthetic field VHE is driven solely by the skew scattering,  and one can expect temperature-dependent VH current since  $\chi \propto {T}$ in Eq.~\eqref{i:vhe:skew:2ph}. 

{To estimate the effect and link with electronic VHE we introduce}
VH `conductivity' ${\sigma_{\rm VH} = }e^2\chi N$ and take into account that the factors $\mathcal A \sim 1$. 
For anomalous and coherent skew scattering contributions ${\sigma_{\rm VH} ^{(anom)}} \sim (e^2/\hbar) \xi N$. Using generic estimates $\xi \sim 10\ldots 100$~\AA$^2$~\cite{Xiao:2012cr}, $N \sim 10^{12}$~cm$^2$~\cite{note3}
the VH conductivity ${\sigma_{\rm VH}^{(anom)}}\sim (10^{-4}\ldots 10^{-3})(e^2/\hbar)$. The contribution~\eqref{i:vhe:skew:2ph} is parametrically different, ${\sigma_{\rm VH} ^{(2ph)}} \sim (10^{-4}\ldots 10^{-3}) \beta(T)(e^2/\hbar)$ and contains the temperature dependent factor $\beta(T) \sim \tilde \Xi_v k_B T/(\Xi_c - \Xi_v)^2$. For  typical values of deformation potentials in the range of units to tens of eV~\cite{PhysRevB.85.115317,PhysRevB.90.045422,PhysRevLett.124.166802} $\beta(T) \ll 1$, but this contribution has a specific temperature dependence. {If the skew scattering is provided by static impurities, corresponding $\sigma_{\rm VH}^{(imp)} \sim \sigma_{\rm VH}^{(anom)} k_B T \tau_p MU/(2\pi \hbar^3)$ where $U$ is the scattering potential Fourier-component. For relatively strong scatterers the factor $M|U|/(2\pi \hbar^2) \sim 1$,  and at $k_B T \tau_p/\hbar \gg 1$ the impurity-induced skew effect can be dominant~\cite{note:suppl,2020arXiv200608717K}.}

\begin{figure}[h]
\includegraphics[width=\linewidth]{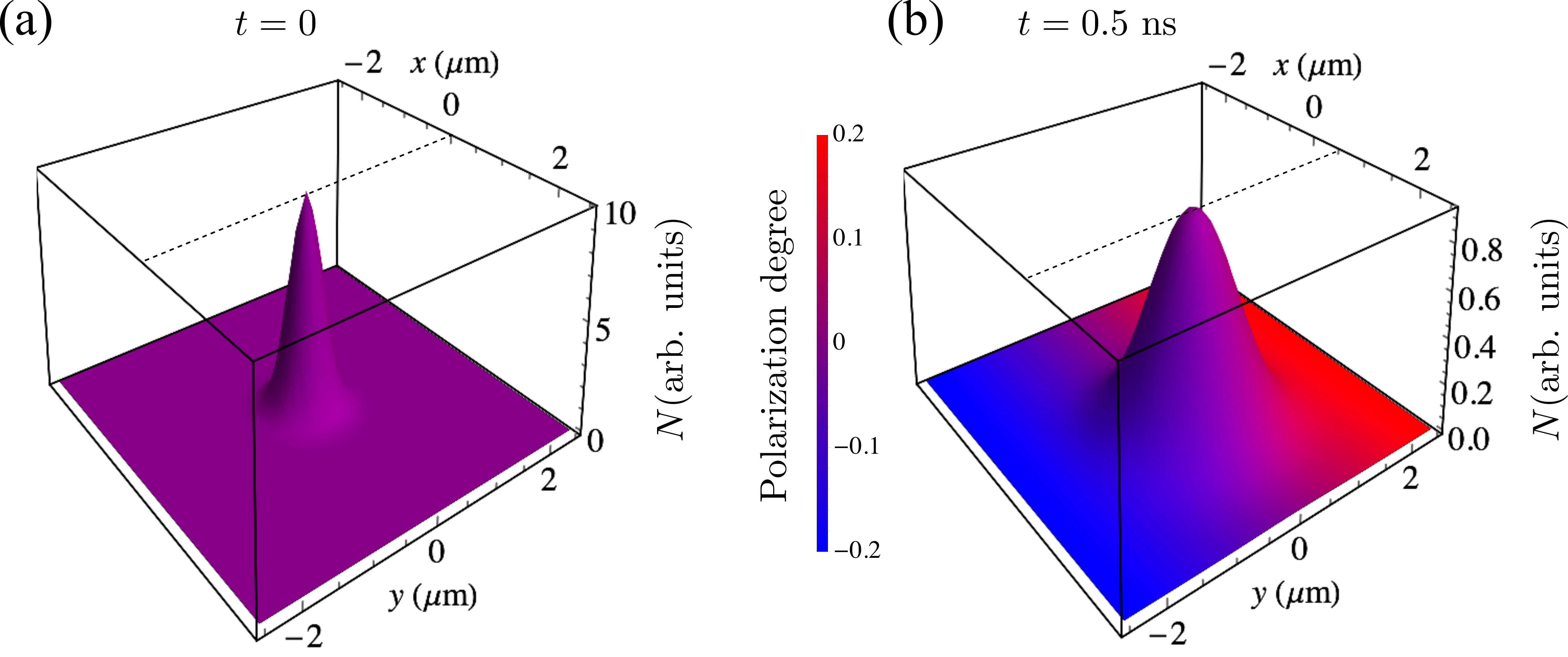}
\caption{Exciton density profile at $t=0$ [panel (a)] and $t=0.5$~ns [panel (b)] calculated after Eqs.~\eqref{drift:diff} in the presence of $\bm F_{d} \parallel x$. Color shows exciton valley polarization degree $P_v = 2S_z/N$. Parameters: $D=3$~cm$^2/$s, $\tau_0=1$~ns, $\tau_s=0.3$~ns, drift velocity $F_d \tau_p/M=1$~$\mu$m/ns, VH angle $\beta=2\chi M/\tau_p=0.1$, initial spot size $0.33$~$\mu$m.} \label{fig:prop}
\end{figure}

\emph{Valley separation of diffusing excitons.} Unlike free charge carriers, the exciton transport is usually monitored optically via the photoluminescence emission with temporal and spatial resolution~\cite{PhysRevLett.120.207401,zipfel2019exciton,Rivera688,Lundt:2019aa,Unuchek:2019aa} and the valley polarization of excitons is mapped to
the circular polarization of the radiation. Extending the theory of Refs.~\cite{dyakonov_book,PhysRevLett.120.207401,PhysRevB.100.045426} we formulate the set of drift-diffusion equations for the exciton density $N(\bm r,t) = N_+(\bm r,t) + N_-(\bm r,t)$ and pseudospin component $S_z(\bm r,t) = [N_+(\bm r,t) - N_-(\bm r,t)]/2$, with $N_\pm$ being the densities of excitons in the $\bm K_\pm$ valleys. The analysis~\cite{note:suppl} shows that
\begin{align}
&\hspace{-0.4cm}\frac{\partial N}{\partial t} = 
D\Delta N - \frac{\tau_p}{M} \bm F_d \cdot \bm \nabla N - 4\chi   [\bm F_{d} \times \bm \nabla]_z S_z - \frac{N}{\tau_0},\label{drift:diff}\\
&\frac{\partial S_z}{\partial t} = D\Delta S_{z} - \frac{\tau_p}{M} \bm F_d \cdot \bm \nabla S_z - \chi  [\bm F_{d} \times \bm \nabla N]_z -  \frac{S_z}{\tau_s},\nonumber
\end{align}
where $\tau_{0}$ is the exciton lifetime, $\tau_s$ is the {valley} polarization lifetime. 
Note that the value of the parameter $\chi$ depends, generally, on the mechanism behind the drag force $\bm F_d$ as shown above~\cite{note4}.
The product $2\chi M/\tau_p$ 
is VH angle because it describes the rotation of the flux of valley-polarized excitons due to the 
VHE. {The set~\eqref{drift:diff} makes it possible to address various regimes of exciton transport. We briefly consider three situations. (i) Excitons diffuse in the absence of external fields and temperature gradients. In this case Eqs.~\eqref{drift:diff} reduce to the diffusion equations for spin and valley polarization and no valley separation of excitons occurs in the bulk of the sample~\cite{note:suppl}. (ii) Excitons are generated homogeneously and the drag force is applied. In this case the exciton valley polarization $\propto \pm \chi[\hat{\bm z}\times \bm F_d]$ is generated in the vicinity of the edges with the spatial scale $\sqrt{D\tau_s}$~\cite{note:suppl}. (iii) The homogeneous drag force $\bm F_d\parallel x$ is present, but excitons are generated at $t=0$ by a focused spot. In this case, the valley separation occurs in the drifting exciton cloud.} The exciton and exciton valley polarization degree spatial distributions calculated by solving Eqs.~\eqref{drift:diff} are presented in Fig.~\ref{fig:prop}.  
The exciton cloud spreads due to the diffusion and drifts along the $x$-axis due to the drag force. Importantly, the valley polarization distribution (shown by color) arises due to the VHE. The exciton polarization degree is an odd function of the transversal to the $\bm F_d$ coordinate, $y$~\cite{note:suppl}.

\emph{Conclusion.} To conclude, we have developed the microscopic theory of the exciton VHE in two-dimensional transition metal dichalcogenides. The theory takes into account all relevant contributions to the VH current. We demonstrate the cancellation of the anomalous velocity effect by the part of the side-jump contribution and stress that this cancellation is general and does not depend on the specifics of the model. We show that the resulting VH current is driven by the side-jump and skew scattering mechanisms and calculate microscopically these contributions. The resulting VH  current depends crucially both on the details of the scattering and on the origin of the drag force, i.e., whether the excitons drift due to a synthetic field or due to the phonon drag. 
We include the VHE in the drift-diffusion description of the exciton transport and illustrate the valley separation of the excitons due to the VHE. 

\acknowledgements 
The reported study was funded by RFBR and CNRS, project number 20-52-16303. The work of L.E.G. was supported by the Foundation for the Advancement of Theoretical Physics and Mathematics ``BASIS''.


\begin{thebibliography}{58}
\bibitem{note:static:linear} {Strictly speaking, Eq.~\eqref{iVHE} holds in $\bm F_d$-linear regime and assuming that $\omega\tau_p \ll 1$, where $\omega$ is the characteristic frequency of the force variation and $\tau_p$ is the exciton momentum relaxation time.}
\bibitem{note:full:strain} {We neglect valley-dependent interband contributions due to the shear strain components assuming $\epsilon_{xx}=\epsilon_{yy}$ and $\epsilon_{xy}=\epsilon_{yx}=0$, see, e.g.,~\cite{PhysRevLett.113.077201}.}
\bibitem{note1} These contributions are partially responsible for the side-jump~\cite{nozieresAHE,dyakonov_book}.
\bibitem{note2} The situation of exciton-impurity scattering is presented in SI~\cite{note:suppl}.
\bibitem{note3} This density is significantly  below the Mott density due to the high binding energy of the exciton.
\end{thebibliography}
\end{document}


\title{
Online supplementary information:\\
Skew Scattering and Side Jump Drive Exciton Valley Hall Effect\\ in Two-Dimensional Crystals
}

\author{M.~M.~Glazov}
\author{L.~E.~Golub} 	
\affiliation{Ioffe Institute,  	194021 St.~Petersburg, Russia}

\maketitle

\tableofcontents

\section{Exciton Bloch functions and Berry curvature}

Here we demonstrate the way of construction of excitonic Bloch functions within the three-band $\bm k\cdot \bm p$-model. We start with the Hamiltonians $\mathcal H_\pm$ describing the electron states (in the electron representation) in the vicinity of the $\bm K_\pm$ points of the Brillouin zone:
\begin{equation}
\label{Hpm}
\mathcal H_\pm = \begin{pmatrix}
0 & \pm \gamma k_\mp & \pm\gamma' k_\pm\\
\pm\gamma k_\pm & - E_g & 0 \\
\pm\gamma' k_\mp & 0 & -E_g'
\end{pmatrix}
\end{equation}
Here $k_\pm = k_x \pm\mathrm i k_y$, the wavevector is reckoned from the corresponding Brillouin zone edge, real constants $\gamma$ and $\gamma'$ are the parameters related to the interband momentum matrix elements, $E_g$ is the band gap and $E_g'$ is the energy distance from the conduction band to the remote valence band, $\bm k\cdot \bm p$ mixing between the valence band and the $k^2$ diagonal terms are disregarded for brevity. Note that the states in $\bm K_+$ and $\bm K_-$ valleys are related by the time-reversal operation and the relation between $\mathcal H_+$ and $\mathcal H_-$ establishes the rules of the transformation at the time reversal (i.e., the phases of the Bloch states):
\begin{equation}
\label{tr}
k_\pm \to -k_\mp.
\end{equation}
Accordingly, in the lowest non-vanishing order in the $\bm k\cdot \bm p$-mixing the electron Bloch function acquires the form~\cite{2020arXiv200405091G}
\begin{equation}
\label{e:Bloch}
|e,\bm k, \pm\rangle \propto |c,\pm\rangle \pm\frac{\gamma}{E_g} k_\pm |v,\pm\rangle \pm  \frac{\gamma'}{E_g'} k_\pm |v',\pm\rangle.
\end{equation}
Here $|c,\pm\rangle$, $|v,\pm\rangle$, and $|v',\pm\rangle$ are the Bloch functions at the $\bm K_\pm$ points of the Brillouin zone, the normalization constant is omitted. Equation~\eqref{e:Bloch} is in agreement with Eq.~(3a) of the main text.

In the course of direct optical transition the electron is promoted from the valence band to the conduction band leaving behind an unoccupied state in the same valley. According to the general theory~\cite{birpikus_eng,ivchenko05a,PSSB:PSSB201552211,PhysRevB.95.035311,RevModPhys.90.021001}, the hole state is associated with the state, obtained by the time reversal operation of the unoccupied state (in this way the linear and angular momentum conservation laws at optical transitions are naturally fulfilled, see also the the theory of positronium~\cite{ll4_eng}). Let us for specificity consider the electron-hole pair excited by the $\sigma^+$ light, i.e., where both the electron and unoccupied state are in the $\bm K_+$ valley, correspondingly, the hole is in the $\bm K_-$ valley. The matrix elements ($n,n'$ are the band indices, $\mathcal K$ is the time reversal operator, superscript $(h)$ underlines the hole representation) of the hole Hamiltonian are obtained as~\cite{birpikus_eng}:
\begin{equation}
\label{H:hole:gen}
\mathcal H_{nn'}^{(h)}(\bm k) = -\mathcal H_{\mathcal Kn',\mathcal Kn}(-\bm k),
\end{equation}
with the result
\begin{equation}
\label{H:hole:-}
\mathcal H_{-}^{(h)}(\bm k) = \begin{pmatrix}
0 & \pm \gamma k_\mp & \pm\gamma' k_\pm\\
\pm\gamma k_\pm & + E_g & 0 \\
\pm\gamma' k_\mp & 0 & +E_g'
\end{pmatrix}.
\end{equation}
For completeness we note that the Hamiltonian for the hole in the $\bm K_+$ valley (i.e., the hole excited by the $\sigma^-$ light, unoccupied state is in the $\bm K_-$ valley) can be obtained from~\eqref{H:hole:-} by the transformation Eq.~\eqref{tr}.
The hole state in the lowest order in the $\bm k\cdot \bm p$-mixing reads in agreement with Eq. (3b) of the main text
\begin{equation}
\label{h:Bloch}
|h,\bm k, \mp \rangle \propto |\tilde v,\mp\rangle \mp\frac{\gamma}{E_g} k_\pm |\tilde c,\mp\rangle ,
\end{equation}
where tilde denotes the hole Bloch functions.

In what follows we assume that the Coulomb interaction is sufficiently weak such that the exciton binding energy $E_B$ is by far smaller than the band gaps $E_g$, $E_g'$. Thus, the Coulomb interaction can be considered as a perturbation and the exciton state can be presented as a linear superposition of the two-particle Bloch functions~\cite{birpikus_eng,RevModPhys.90.021001}
\begin{equation}
\label{exc:gen}
\Psi_{exc,\pm} = \sum_{\bm k_e, \bm k_h} e^{\mathrm i \bm k_e\bm r_h+ \mathrm i \bm k_h \bm r_h} \Phi(\bm k_e, \bm k_h) |e,\bm k_e, \pm \rangle|h,\bm k_h, \mp \rangle.
\end{equation}
Here subscripts $\pm$ denote the polarization of light which excites corresponding exciton, $\Phi(\bm k_e, \bm k_h)$ is the smooth envelope of the exciton wavefunction in the $\bm k$-space, $\bm k_e$, $\bm k_h$ are the electron and hole wavevectors. The envelope function can be found from the two-particle Schr\"odinger equation written in the effective mass approximation. For a free exciton the total wavevector $\bm K=\bm k_e+\bm k_h$ is a good quantum number and
\begin{equation}
\label{env}
\Phi(\bm k_e, \bm k_h) = \delta_{\bm k_e+\bm k_h,\bm K} F(\bm k),
\end{equation}
with 
\[
F(\bm k) =\int d\bm \rho \varphi(\bm \rho) e^{-\mathrm i \bm k\bm \rho} d\bm \rho,
\]
is the relative motion envelope function with $\varphi(\bm \rho)$ being the real space envelope function. For completeness we present also the relations between the coordinates (and wavevectors) of the individual carriers, $\bm r_e$ and $\bm r_h$ ($\bm k_e$, $\bm k_h$) and center-of-mass/relative motion:
\begin{subequations}
\label{Rrho}
\begin{align}
\bm R = \frac{m_e}{M} \bm r_e + \frac{m_h}{M}\bm r_h,\quad \bm \rho = \bm r_e - \bm r_h,\\
\bm k_e = \frac{m_e}{M}\bm K + \bm k,\quad
\bm k_h = \frac{m_h}{M} \bm K - \bm k,
\end{align}
\end{subequations}
with $m_e$, $m_h$ and $M=m_e+m_h$ being the electron, hole and exciton mass, respectively.

Equations~\eqref{exc:gen} and \eqref{env} are in agreement with Eq. (2) of the main text:
\begin{equation}
\label{exciton:bloch}
\Psi_{exc,\pm} = e^{\mathrm i \bm K \bm R} \mathcal U_{\bm K,\pm}(\bm \rho),
\end{equation}
where the function
\begin{equation}
\label{amplitude:exc}
\mathcal U_{\bm K,\pm}(\bm \rho) = \sum_{\bm k} e^{\mathrm i \bm k\bm \rho} F(\bm k) |e,\bm k_e,\pm \rangle|h,\bm k_h,\mp \rangle,
\end{equation}
can be formally associated with the exciton Bloch amplitude.
Note that the optically active excitons have the center of mass wavevectors in the vicinity of the Brillouin zone center ($\Gamma$-point). This is because electron and hole occupy opposite valleys of the $\bm k$-space. Making use of definition of the exciton Bloch amplitude~\eqref{amplitude:exc} one can express the excitonic Berry curvature $z$-component as
\begin{equation}
\label{exc:Berry:z}
\mathcal F_{\pm,z} =  { 2\Im \int  \frac{\partial \mathcal U_{\bm K,\pm}^\dag(\bm \rho)}{\partial K_y} \frac{\partial \mathcal U_{\bm K,\pm}(\bm \rho)}{\partial K_x}  d\bm \rho } = \mp  2\left[\xi_e\left(\frac{m_e}{M}\right)^2  + \xi_h \left(\frac{m_h}{M}\right)^2\right]\hat{\bm z}.
\end{equation}
Here, within the considered three-band model
\begin{subequations}
\label{xi:eh}
\begin{align}
\xi_e = \frac{\gamma^2}{E_g^2} + \frac{\gamma'^2}{E_g'^2},\\
\xi_h = \frac{\gamma^2}{E_g^2},
\end{align}
\end{subequations}
and Eq.~\eqref{exc:Berry:z} corresponds to Eq.~(4) of the main text.

Making use of Eq.~\eqref{amplitude:exc} one can readily calculate the contribution to the  exciton position operator related to the Bloch bands:
\begin{equation}
\label{exc:pos}
\bm \Omega_{\pm, \bm K} = {-\Im \int \mathcal U_{\bm K,\pm}^\dag(\bm \rho)\bm \nabla_{\bm K} \mathcal U_{\bm K,\pm}(\bm \rho) d\bm\rho}= \frac{1}{2}[\bm{\mathcal F}_\pm \times \bm K],
\end{equation}
in agreement with Eq.~(5) of the main text. In fact, one can also arrive to Eq.~\eqref{exc:pos} using simple qualitative arguments evaluating the center-of-mass coordinate of the exciton:
\begin{equation}
\label{exc:pos:simple}
\bm\Omega_{\pm ,\bm K} = \overline{\frac{m_e}{M} \bm \Omega_{\pm,e}(\bm k_e) + \frac{m_h}{M} \bm \Omega_{\mp,h}(\bm k_h)},
\end{equation}
where $\bm \Omega_{\pm, e}$, $\bm \Omega_{\pm,h}$ are the electron and hole Bloch coordinates and the overline denotes averaging over the relative motion wavevector $\bm k$.

{The analysis performed above is based on the assumptions of small exciton binding energy $E_B \ll E_g,E_g'$ and small kinetic energy, $\hbar^2 K^2/2M \ll E_g, E_g'$. While the kinetic energy of the excitons is small even at the room temperature, the binding energy depending on the dielectric environment of the monolayer can be significant and reach roughly a quarter of the band gap~\cite{RevModPhys.90.021001}. This calls for advanced description of the excitonic states in terms of the multicomponent envelope functions (see, e.g.,~\cite{PhysRevLett.120.187401,PhysRevB.96.035131,2020arXiv200704839L} and references therein), or numerical approaches based on the direct solution of Bethe-Salpeter equation along the lines of Ref.~\cite{PhysRevB.91.075310}. Altogether, it results in a more complex form of the exciton Bloch functions \eqref{exciton:bloch} and \eqref{amplitude:exc} and renormalization of the Berry curvature $\mathcal F_{\pm,z}$ and coordinate $\bm \Omega_{\pm,\bm K}$ (resulting, e.g., from the higher-order $\bm k\cdot \bm p$ terms in Eqs.~\eqref{e:Bloch} and \eqref{h:Bloch}), as well as the scattering matrix element. Derivation of these quantities is beyond the scope of the paper. Here we just mention that, as it follows from the symmetry analysis and general arguments, for small exciton kinetic energy ($K\to 0$) $\mathcal F_{\pm,z}$ is a constant and $\bm\Omega_{\pm,\bm K}$ is still given by the right hand side of Eq.~\eqref{exc:pos}. }

\section{Exciton scattering by external potential}

We introduce the perturbation operator $\mathcal V$ related to the external potential (i.e., due to the defects of the crystal or the deformation potential due to phonons) as
\begin{equation}
\label{V:electron}
\mathcal V = \begin{pmatrix}
V_c(\bm r) & 0 & 0\\
0 & V_v(\bm r) & 0 \\
0 & 0 & V_{v'}(\bm r)
\end{pmatrix}.
\end{equation}
For simplicity we assume that the perturbation is diagonal in the band indices and does not mix valleys, which is reasonable assumption for the long-wavelength acoustic phonon scattering. {Strictly speaking, the shear strain with $\epsilon_{xy} \ne 0$ or $\epsilon_{xx} \ne \epsilon_{yy}$ induces the valley-dependent interband coupling, cf. Ref.~\cite{PhysRevLett.113.077201}. It is controlled by independent parameters and we neglect this coupling hereafter.} The perturbation $\mathcal V$ is presented in the electron representation. It follows from Eq.~\eqref{H:hole:gen} that in the hole representation the perturbation has the same form, but the opposite sign. 

Making use of  definitions of the electron and hole Bloch functions, Eqs.~\eqref{e:Bloch} and \eqref{h:Bloch}, we arrive at the following expressions for the scattering matrix elements for the electron in the $\bm K_+$ valley and the hole in the $\bm K_-$ valley 
\begin{subequations}
\label{scatter}
\begin{equation}
\label{e:scatter}
V^e_{\bm k_e',\bm k_e} = V_c(\bm k_e' - \bm k_e) + \mathrm i \xi [\bm k_e' \times \bm k_e]_z V_v(\bm k_e' - \bm k_e) - \mathrm i \xi' [\bm k_e' \times \bm k_e]_z V_{v'}(\bm k_e' - \bm k_e), 
\end{equation}
\begin{equation}
\label{h:scatter}
V^h_{\bm k_e',\bm k_e} =-V_v(\bm k_e' - \bm k_e) - \mathrm i \xi [\bm k_h' \times \bm k_h]_z V_c(\bm k_e' - \bm k_e),
\end{equation}
\end{subequations}
with 
\begin{equation}
\label{xis}
\xi = \gamma^2/E_g^2,\qquad \xi' = \gamma'^2/E_g'^2.
\end{equation} 
{Hereafter we neglect valley independent contributions to the matrix elements $\propto \xi(\bm k\cdot \bm k'), \xi'(\bm k\cdot \bm k')$. These contributions do not play a role in the exciton valley Hall effect in the lowest order in its kinetic energy.}
Equations~\eqref{scatter} together with~\eqref{exciton:bloch} allow us to express the exciton scattering matrix element in the form
\[
M_{\bm K',\bm K} = \sum_{\substack{\bm k_e,\bm k_h\\\bm k_e',\bm k_h'}}\delta_{\bm K, \bm k_e+ \bm k_h} \delta_{\bm K',\bm k_e'+\bm k_h'} F^*(\bm k') F(\bm k)
\left( V^e_{\bm k_e',\bm k_e} \delta_{\bm k_h,\bm k_h'} +V^h_{\bm k_h',\bm k_h} \delta_{\bm k_e,\bm k_e'}\right){.}
\]
{In what follows we assume that the translational motion wavevectors of exciton in the initial and final states $\bm K$ and $\bm K'$ are by far smaller than the inverse Bohr radius $a_B$. Thus, taking into account that $\bm k' = (\bm K' - \bm K)m_h/M + \bm k$ and neglecting the contributions $\propto (\bm K' - \bm K)$ in the argument of $F(\bm k')$, also making use of the expression
\[
\bm k_e - \bm k_e' = \bm K \frac{m_e}{M} + \bm k - \bm K' \frac{m_e}{M} -\bm k' = \bm K - \bm K',
\]
and analogous one for the holes,
we arrive at
}
\begin{multline}
\label{exc:scatter}
M_{\bm K',\bm K} = V_c(\bm K' - \bm K) - V_{v} (\bm K' - \bm K) + \mathrm i \xi V_{v} (\bm K' - \bm K) \frac{m_e}{M} [\bm K' \times \bm K]_z\\
 + \mathrm i \xi' V_{v'} (\bm K' - \bm K) \frac{m_e}{M} [\bm K' \times \bm K]_z
 - \mathrm i \xi V_c(\bm K' - \bm K) \frac{m_h}{M} [\bm K' \times \bm K].
\end{multline}
Note that for the exciton created by the $\sigma^-$ light where the electron occupies the $\bm K_-$ valley and the hole occupies the $\bm K_+$ valley the signs of $\xi$ and $\xi'$ should be reversed.

It is instructive to disregard the contribution $\xi'$ to the Berry curvature and scattering matrix element. In this case in agreement with the main text [Eq.~(6)] we obtain from Eq.~\eqref{exc:scatter}
\begin{equation}
\label{scatt}
M_{\bm K'\bm K} = V_c(\bm K' - \bm K) - V_v(\bm K' - \bm K)
{- \mathrm i \xi \left[V_c(\bm K' - \bm K)\frac{m_h}{M} - V_v(\bm K' - \bm K) \frac{m_e}{M} \right][\bm K'\times \bm K]_z.}
\end{equation}
Equation~\eqref{scatt} can be further simplified in the relevant case of the short range scattering {where the potential correlation length $l_c$ is much smaller than $1/K,1/K'$ and $V_c$, $V_v$ can be taken as wavevector independent. For example, for acoustic phonon scattering via the deformation potential mechanism this requires that $Ka_B, K'a_B \ll 1$, the condition fulfilled in a wide range of temperatures~\cite{shree2018exciton}. Also, we disregard the transitions from the exciton ground state to the excited states of relative motion of electron-hole pair in the exciton since these processes are forbidden by the energy conservation at $\hbar^2 K^2/(2M) \ll E_B$}. We introduce the notations 
\begin{equation}
\label{not}
\nu = \frac{V_c}{V_v}, \qquad \mathcal A  =\frac{\nu{m_h} - {m_e}}{M({\nu - 1})},
\end{equation}
 and recast 
\begin{equation}
\label{exc:scatt:simpl}
M_{\bm K'\bm K}= (V_c - V_v)\left(1- \mathrm i \xi {\cal A}  [\bm K' \times \bm K]_z\right).
\end{equation}
Note that the analogous scattering matrix element for electrons takes the form
\begin{equation}
\label{e:scatt:simpl}
M_{\bm k_e'\bm k_e}=V_c\left(1+ \mathrm i {\xi\over \nu} [\bm k_e'\times \bm k_e]_z\right),
\end{equation}
i.e., it differs from Eq.~\eqref{exc:scatt:simpl} by the replacement 
\begin{equation}
\label{repl}
\nu^{-1} \to - \mathcal A.
\end{equation}
 It makes possible to use the results of Ref.~\cite{2020arXiv200405091G} for calculation of the exciton valley Hall effect.

\section{Anomalous contributions to the excitonic VHE}

It is shown in Ref.~\cite{2020arXiv200405091G} that for two-dimensional electrons in the presence of a force $\bm F$ the side-jump contribution due to the change of potential energy at scattering, $e\bm i_{v}^{(sj,b)}$, is given by 
\begin{equation}
\label{sj:electrons:F}
e\bm i_{v}^{(sj,b)} = e\bm i_{v}^{(sj,a)}= -{1\over 2}eN_e[\bm{\mathcal F}\times \bm F] + {e\over {\hbar}}  N_e{\xi\over \nu}[\hat{\bm z} \times \bm F].
\end{equation}
Here $N_e$ is the electron density per valley, and $\bm i_{v}^{(sj,a)}$ is the side-jump accumulation contribution, cf. Eq.~(12) of the main text. Since the form of the scattering matrix elements for electrons and excitons is similar, cf. Eqs.~\eqref{exc:scatt:simpl} and~\eqref{e:scatt:simpl}, the corresponding side-jump current for excitons in the presence of \emph{synthetic field} can be derived from Eq.~\eqref{sj:electrons:F} using the replacement~\eqref{repl} together with  obvious substitutions (we recall that $N$ is the total density of excitons):
\begin{equation}
\label{subst}
{\bm F \to \bm F_d^{(s)} }, \qquad N_e \to \frac{1}{2} N.
\end{equation}
As a result, we obtain
for the total side-jump current ${\bm i_{v}^{(sj)}=\bm i_{v}^{(sj,a)}+\bm i_{v}^{(sj,b)} }$ Eq. (13) of the main text:
\begin{equation}
\label{i:vhe:sj}
\bm i_{v}^{(sj)} = -\frac{N}{2} [\bm{\mathcal F}\times \bm F_d^{(s)}] -  {\frac{\xi}{\hbar}\mathcal A} N [\hat{\bm z} \times \bm F_d^{(s)}].
\end{equation}
The first term here exactly compensates the anomalous velocity contribution, and the total anomalous current is given by the linear in ${\cal A}$ part of the side-jump contribution. {It is noteworthy that the anomalous velocity and side-jump contributions are determined, in our approximations, by the total density of the particles, while the form of their distribution function is unimportant provided that the energy relaxation is slow as compared with the momentum relaxation. That is why the compensation holds for any arbitrary  (non-equilibrium) distribution function of excitons but in the linear in $\bm F_d$-regime.}
At the phonon drag, the anomalous contribution to the electron VHE has the form~\cite{2020arXiv200405091G}
\begin{equation}
\chi^{(anom,d)}_{el} = \frac{\xi}{\hbar} \left({1\over \nu_{\rm imp}} - {1\over \nu_{\rm ph}}\right) \frac{\tau_p}{\tau_{\rm imp}}.
\end{equation}
Making the substitutions~\eqref{repl} and \eqref{subst} we get the expression for excitons:
\begin{equation}
\label{i:vhe:anom:drag}
\chi^{(anom,d)} = -\frac{\xi}{2\hbar} (\mathcal A_{\rm imp} - \mathcal A_{\rm ph}) \frac{\tau_p}{\tau_{\rm imp}},
\end{equation}
in agreement with Eq. (16) of the main text.

\section{Skew scattering contributions}

\subsection{Exciton-impurity scattering}

For non-degenerate electrons scattered by short-range impurities, the skew scattering contribution to VHE in the electric field is given by~\cite{2020arXiv200405091G}
\begin{equation}
\chi^{(skew,imp)}_{el} = 4\pi {\xi \over \hbar}gU_v {k_BT\tau_p\over\hbar}
= 2{\xi \over \hbar} {k_BT \over \nu n_i U_c}.
\end{equation}
Here $U_{c,v}$ are the Fourier components of the scattering
potential acting in the conduction and valence bands, cf. Eq.~\eqref{V:electron}, and $n_i$ is the impurity density.
Making the substitutions~\eqref{repl} and~\eqref{subst} we obtain for excitons
\begin{equation}
\chi^{(skew,imp)}= -{\cal A}{\xi \over \hbar} {k_BT \over  n_i U_c}.
\end{equation}

\subsection{Two-phonon scattering}

Let us firstly analyze the contributions due to the two-phonon processes described by the Hamiltonian~(17) of the main text. Analogous calculation for non-degenerate electrons yields~\cite{2020arXiv200405091G}:
\begin{equation}
\chi^{(skew,2ph,s)}_{el} = -4{\frac{\xi}{\hbar}}  k_B T {\tilde{\Xi}_v + 2{\Xi_v\over \Xi_c}\tilde \Xi_c\over \Xi_c^2},
\end{equation}
where $\Xi_{c,v}$ and $\tilde \Xi_{c,v}$ are the one- and two-phonon constants for the conduction and valence bands.
For excitons, we make the following substitutions in accordance with Eq.~\eqref{scatt}
\begin{align}
\Xi_c \to \Xi_c-\Xi_v, \quad \Xi_v \to {m_e\Xi_v - m_h\Xi_c\over M}, \nonumber \\
\tilde \Xi_c \to \tilde \Xi_c-\tilde \Xi_v, \quad \tilde \Xi_v \to {m_e\tilde \Xi_v - m_h\tilde \Xi_c\over M}.
\end{align}
Therefore we obtain Eq. (18) of the main text:
\begin{equation}
\label{i:vhe:skew:2ph}
\chi^{(skew,2ph,s)} = {\frac{\xi}{\hbar}}  {\mathcal B} \frac{ k_B T\tilde \Xi_v}{(\Xi_c - \Xi_v)^2 } ,
\end{equation}
where
\[
\mathcal B =  2\mathcal A_{\rm ph} \left[\frac{m_e}{M} - \frac{m_h}{M}\tilde \nu_{\rm ph} + 2\mathcal A_{\rm ph}(1-\tilde \nu_{\rm ph})\right],
\qquad \tilde \nu_{\rm ph} = \tilde \Xi_c/\tilde \Xi_v.
\]
In a similar fashion Eqs. (19) and (20) of the main text can be derived from the results of Ref.~\cite{2020arXiv200405091G}.

%
%
%
%

\subsection{Coherent skew scattering at the phonon drag}

Let us now turn to the coherent two-phonon skew scattering processes in the case where the scattering takes place on the anisotropic distribution of the phonons. For clarity and brevity of presentation we consider the case electrons scattered by the anisotropic distribution of phonons and calculate the asymmetric scattering in $\bm K_+$ valley as follows:
\begin{multline}
W^{(as)}_{\bm k' \bm k} = {2\pi\over \hbar} \sum_{\bm p\bm p',\nu,\mu = \pm}
\delta(\varepsilon_{k'}-\varepsilon_{k}-\nu\hbar s |\bm k-\bm p|+\mu\hbar s |\bm k'-\bm p|)\delta_{\bm k+\bm k',\bm p+ \bm p'} \\
\times {1\over 2}\left|{M^{(2,\mu)}_{\bm k' \bm p}M^{(1,-\nu)}_{\bm p \bm k}\over \varepsilon_k - \varepsilon_p + \nu\hbar s|\bm k  -\bm p| + i0} +{M^{(1,-\nu)}_{\bm k' \bm p'}M^{(2,\mu)}_{\bm p' \bm k}\over \varepsilon_k - \varepsilon_{p'} - \mu\hbar s|\bm k  -\bm p'| + i0} \right|_{as}^2.
\end{multline}
Here the factor $1/2$ allows avoiding double account for the same phonons. Taking the interference term we obtain
\begin{multline}
\text{Re}{M^{(2,\mu)}_{\bm k' \bm p}M^{(1,-\nu)}_{\bm p \bm k}\left[M^{(1,-\nu)}_{\bm k' \bm p'}M^{(2,\mu)}_{\bm p' \bm k}\right]^*\over (\varepsilon_k - \varepsilon_p + \nu\hbar s|\bm k  -\bm p| + i0)(\varepsilon_k - \varepsilon_{p'}- \mu\hbar s|\bm k  -\bm p'| - i0)}\\
\to \pi \text{Im}\left\{M^{(2,\mu)}_{\bm k' \bm p}M^{(1,-\nu)}_{\bm p \bm k}\left[M^{(1,-\nu)}_{\bm k' \bm p'}M^{(2,\mu)}_{\bm p' \bm k}\right]^*\right\}
\left[
{\delta(\varepsilon_k - \varepsilon_p + \nu\hbar s|\bm k  -\bm p|) \over \varepsilon_k - \varepsilon_{p'}- \mu\hbar s|\bm k  -\bm p'|}
-{\delta(\varepsilon_k - \varepsilon_{p'}- \mu\hbar s|\bm k  -\bm p'|) \over \varepsilon_k - \varepsilon_p + \nu\hbar s|\bm k  -\bm p|}
\right]
\\
= 2 \pi {\delta(\varepsilon_k - \varepsilon_p + \nu\hbar s|\bm k  -\bm p|) \over \varepsilon_k - \varepsilon_{p'}- \mu\hbar s|\bm k  -\bm p'|}\text{Im}\left\{M^{(2,\mu)}_{\bm k' \bm p}M^{(1,-\nu)}_{\bm p \bm k}\left[M^{(1,-\nu)}_{\bm k' \bm p'}M^{(2,\mu)}_{\bm p' \bm k}\right]^*\right\},
\end{multline}
where we used that at simultaneous change $\bm p \leftrightarrow \bm p'$, $\mu \leftrightarrow -\nu$ the product of matrix elements is changed to the complex-conjugated.

Using the property $\left[M^{(\nu)}_{\bm k \bm k'}\right]^* = M^{(-\nu)}_{\bm k' \bm k}$,
we obtain
\begin{multline}
\label{W_as}
W^{(as)}_{\bm k' \bm k} = {(2\pi)^2\over \hbar} \sum_{\bm p\bm p',\nu,\mu = \pm}
\delta(\varepsilon_{k'}-\varepsilon_{k}-\nu\hbar s |\bm k-\bm p|+\mu\hbar s |\bm k'-\bm p|)\delta_{\bm k+\bm k',\bm p+ \bm p'} \\
\times{\delta(\varepsilon_k - \varepsilon_p + \nu\hbar s|\bm k  -\bm p|) \over \varepsilon_k - \varepsilon_{p'}- \mu\hbar s|\bm k  -\bm p'|}\text{Im}
\left\{ \left<M^{(\mu)}_{\bm k' \bm p}M^{(-\mu)}_{ \bm k\bm p'}\right>_{ph} \left<M^{(-\nu)}_{\bm p \bm k}M^{(\nu)}_{\bm p' \bm k'}\right>_{ph}\right\}.
\end{multline}

One can check that the above expression indeed changes sign at substitution $\bm k \leftrightarrow \bm k'$:
\begin{multline}
W^{(as)}_{\bm k \bm k'} = {(2\pi)^2\over \hbar} \sum_{\bm p\bm p',\nu,\mu = \pm}
\delta(\varepsilon_{k}-\varepsilon_{k'}-\nu\hbar s |\bm k'-\bm p|+\mu\hbar s |\bm k-\bm p|)\delta_{\bm k+\bm k',\bm p+ \bm p'} \\
\times{\delta(\varepsilon_{k'} - \varepsilon_p + \nu\hbar s|\bm k'  -\bm p|) \over \varepsilon_{k'} - \varepsilon_{p'}- \mu\hbar s|\bm k'  -\bm p'|}\text{Im}
\left\{ \left<M^{(\mu)}_{\bm k \bm p}M^{(-\mu)}_{ \bm k'\bm p'}\right>_{ph} \left<M^{(-\nu)}_{\bm p \bm k'}M^{(\nu)}_{\bm p' \bm k}\right>_{ph}\right\}
\\ = {(2\pi)^2\over \hbar} \sum_{\bm p\bm p',\nu,\mu = \pm}
\delta(\varepsilon_{k}-\varepsilon_{k'}-\nu\hbar s |\bm k'-\bm p|+\mu\hbar s |\bm k-\bm p|)\delta_{\bm k+\bm k',\bm p+ \bm p'} \\
\times{\delta(\varepsilon_{k} - \varepsilon_p + \mu\hbar s|\bm k  -\bm p|) \over \varepsilon_{k} - \varepsilon_{p'}- \nu\hbar s|\bm k  -\bm p'|}\text{Im}
\left\{ \left<M^{(-\mu)}_{ \bm p\bm k}M^{(\mu)}_{\bm p' \bm k'}\right>^*_{ph} \left<M^{(\nu)}_{ \bm k'\bm p}M^{(-\nu)}_{ \bm k\bm p'}\right>^*_{ph}\right\}.
\end{multline}
If we interchange two dummy indices $\mu \leftrightarrow \nu$ then we see that this expression differs from $W^{(as)}_{\bm k' \bm k}$ just by the complex conjugations which changes the sign of the imaginary part.

In the lowest order in the phonon energy we have
\begin{multline}
\delta(\varepsilon_{k'}-\varepsilon_{k}-\nu\hbar s |\bm k-\bm p|+\mu\hbar s |\bm k'-\bm p|){\delta(\varepsilon_k - \varepsilon_p + \nu\hbar s|\bm k  -\bm p|) \over \varepsilon_k - \varepsilon_{p'}- \mu\hbar s|\bm k  -\bm p'|}  \equiv
{\delta(\varepsilon_{k'}-\varepsilon_{p}+\mu\hbar s |\bm k'-\bm p|)  \over \varepsilon_k - \varepsilon_{p'}- \mu\hbar s|\bm k  -\bm p'|}\delta(\varepsilon_k - \varepsilon_p + \nu\hbar s|\bm k  -\bm p|) \\
\approx 
\mu\hbar s |\bm k'-\bm p| \left[{\delta'(\varepsilon_{k'}-\varepsilon_{p})\over \varepsilon_k - \varepsilon_{p'}} + {\delta(\varepsilon_{k'}-\varepsilon_{p})\over (\varepsilon_k - \varepsilon_{p'})^2} \right]\delta(\varepsilon_k - \varepsilon_p) 
+
\nu\hbar s|\bm k  -\bm p|{\delta(\varepsilon_{k'}-\varepsilon_p)\over \varepsilon_k - \varepsilon_{p'}} \delta'(\varepsilon_k - \varepsilon_p).
\end{multline}
%
The electron-phonon matrix element is given by
\begin{equation}
M^{(\nu)}_{\bm k' \bm k} = i\nu \sqrt{\hbar q\over 2\rho s}\left(\Xi_c + i\xi\Xi_v [\bm k' \times \bm k]_z \right)\sqrt{n_{\nu \bm q}} \delta_{\bm k'-\bm k, \nu \bm q}.
\end{equation}
Then we obtain from Eq.~\eqref{W_as} 
\begin{multline}
\label{sk_inel_1}
W^{(as)}_{\bm k' \bm k} =  {(2\pi)^2\over \hbar} \Xi_c^3 \xi \Xi_v 
 \left({ \hbar\over2\rho s} \right)^2 \hbar s  
 \sum_{\nu, \mu=\pm 1, \bm p, \bm p'} \delta_{\bm k+\bm k',\bm p+ \bm p'}
 {[\bm k' \times \bm p + \bm k \times \bm p' +  \bm p \times \bm k +  \bm p' \times \bm k']_z \over \varepsilon_k - \varepsilon_{p'}}
\\
\times m_{\mu(\bm p- \bm k' )}m_{\nu(\bm p - \bm k)}
 \left\{\mu |\bm k'-\bm p| \left[{\delta'(\varepsilon_{k'}-\varepsilon_{p})\over \varepsilon_k - \varepsilon_{p'}} + {\delta(\varepsilon_{k'}-\varepsilon_{p})\over (\varepsilon_k - \varepsilon_{p'})^2} \right] \delta(\varepsilon_k - \varepsilon_p) +
\nu|\bm k  -\bm p|\delta(\varepsilon_{k'}-\varepsilon_p) \delta'(\varepsilon_k - \varepsilon_p)\right\}
.
\end{multline}
Here $m_{\bm q}=qn_{\bm q}$.

Hereafter we assume the following nonequilibrium phonon distribution 
\begin{equation}
n_{\bm q} = \bar{n}_q (1 + c_{\bm q}), \qquad c_{-\bm q} = -c_{\bm q},
\end{equation}
where $\bar{n}_q=k_\text{B}T/(\hbar s q)$ is the Planck function and $c_{\bm q}$ describes an anisotropic correction responsible for the non-equilibrium phonon flux in the system. 
Then we have from Eq.~\eqref{sk_inel_1}
\begin{multline}
W^{(as)}_{\bm k' \bm k} =  {(2\pi)^2\over \hbar} \Xi_c^3 \xi \Xi_v 
 \left({ \hbar\over2\rho s} \right)^2 \hbar s  \left({k_\text{B}T\over \hbar s} \right)^2
 \sum_{\nu, \mu=\pm 1, \bm p, \bm p'} \delta_{\bm k+\bm k',\bm p+ \bm p'}
 {[\bm k' \times \bm p + \bm k \times \bm p' +  \bm p \times \bm k +  \bm p' \times \bm k']_z \over \varepsilon_k - \varepsilon_{p'}}
\\
\times
(1+c_{\mu(\bm p - \bm k')})(1+c_{\nu(\bm p - \bm k)}) \\
\times
\left\{\mu |\bm k'-\bm p| \left[{\delta'(\varepsilon_{k'}-\varepsilon_{p})\over \varepsilon_k - \varepsilon_{p'}} + {\delta(\varepsilon_{k'}-\varepsilon_{p})\over (\varepsilon_k - \varepsilon_{p'})^2} \right] \delta(\varepsilon_k - \varepsilon_p) +
\nu|\bm k  -\bm p|\delta(\varepsilon_{k'}-\varepsilon_p) \delta'(\varepsilon_k - \varepsilon_p)\right\}
.
\end{multline}
In the lowest order in inelasticity we have 
\begin{equation}
(1+c_{\mu(\bm p - \bm k')})(1+c_{\nu(\bm p - \bm k)}) \approx 1+c_{\mu(\bm p - \bm k')} +c_{\nu(\bm p - \bm k)},
\end{equation}
and since $\sum\limits_{\nu=\pm} \nu c_{\nu\bm q} = 2c_{\bm q}$, we obtain
\begin{multline}
W^{(as)}_{\bm k' \bm k} =  2{(2\pi)^2\over \hbar} \Xi_c^3 \xi \Xi_v 
 \hbar s  \left({k_\text{B}T\over 2\rho s^2} \right)^2
 \sum_{\bm p, \bm p'} \delta_{\bm k+\bm k',\bm p+ \bm p'}
{[\bm k' \times \bm p + \bm k \times \bm p' +  \bm p \times \bm k +  \bm p' \times \bm k']_z \over \varepsilon_k - \varepsilon_{p'}}
\\
\times
\left\{c_{\bm p - \bm k'} |\bm k'-\bm p| \left[{\delta'(\varepsilon_{k'}-\varepsilon_{p})\over \varepsilon_k - \varepsilon_{p'}} + {\delta(\varepsilon_{k'}-\varepsilon_{p})\over (\varepsilon_k - \varepsilon_{p'})^2} \right] \delta(\varepsilon_k - \varepsilon_p) +
c_{\bm p - \bm k}|\bm k  -\bm p|\delta(\varepsilon_{k'}-\varepsilon_p) \delta'(\varepsilon_k - \varepsilon_p)\right\}
.
\end{multline}

Hereafter we assume the following asymmetry of the phonon distribution typical for the phonon drag~\cite{PhysRevB.100.045426,2020arXiv200405091G}:
\begin{equation}
\label{c_q}
c_{\bm q} = {\bm q \over q} \cdot \bm e
\qquad \text{i.e.} \qquad n_{\bm q} = {k_\text{B}T \over \hbar s q} \left(1 + {\bm q \over q} \cdot \bm e\right),
\end{equation}
where $\bm e$ is  the in-plane vector related to the phonon drag force as:
\begin{equation}
\label{e:vect}
\frac{ms}{\tau_{ph}}\bm e = \bm F_\text{drag}
\end{equation} with $m$ being the electron effective mass. 
Then we have
\begin{multline}
W^{(as)}_{\bm k' \bm k} =  2{(2\pi)^2\over \hbar} \Xi_c^3 \xi \Xi_v 
 \hbar s  \left({k_\text{B}T\over 2\rho s^2} \right)^2
 \sum_{\bm p, \bm p'} \delta_{\bm k+\bm k',\bm p+ \bm p'}
[\bm k' \times \bm p + \bm k \times \bm p' +  \bm p \times \bm k +  \bm p' \times \bm k']_z
\\
\times
\left\{(\bm p - \bm k') \left[{\delta'(\varepsilon_{k'}-\varepsilon_{p})\over \varepsilon_k - \varepsilon_{p'}} + {\delta(\varepsilon_{k'}-\varepsilon_{p})\over (\varepsilon_k - \varepsilon_{p'})^2} \right]\delta(\varepsilon_k - \varepsilon_p) +
(\bm p - \bm k) {\delta(\varepsilon_{k'}-\varepsilon_p) \delta'(\varepsilon_k - \varepsilon_p) \over \varepsilon_k - \varepsilon_{p'}}\right\}\cdot \bm e \\
= - {(2\pi)^2\over \hbar} \Xi_c^3 \xi \Xi_v 
 {\hbar s\over 2}  \left({k_\text{B}T\over \rho s^2} \right)^2 g \left[-\delta(\varepsilon_{k'}-\varepsilon_{k})\Phi_0+ \delta'(\varepsilon_{k'}-\varepsilon_{k})(\Phi_1 - \Phi_2) \right]e_x
.
\end{multline}
Here we directed the $x$ axis along $\bm e$, $g$ is the density of states per valley,
\begin{equation}
\Phi_0 = \left< (p_x - k'_x){[\bm k' \times \bm p + \bm k \times \bm p' +  \bm p \times \bm k +  \bm p' \times \bm k']_z \over (\varepsilon_{p'}-\varepsilon_k)^2}\right>_{\varphi_{\bm p}, p=k=k'},
\end{equation}
\begin{equation}
\Phi_1 = \left< (p_x - k'_x){[\bm k' \times \bm p + \bm k \times \bm p' +  \bm p \times \bm k +  \bm p' \times \bm k']_z \over \varepsilon_{p'}-\varepsilon_k}\right>_{\varphi_{\bm p}, p=k},
\end{equation}
\begin{equation}
\Phi_2 = \left< (p_x - k_x){[\bm k' \times \bm p + \bm k \times \bm p' +  \bm p \times \bm k +  \bm p' \times \bm k']_z \over \varepsilon_{p'}-\varepsilon_k}\right>_{\varphi_{\bm p}, p=k'}.
\end{equation}

Introducing the scattering time
\begin{equation}
{1\over \tau_{ph}} = \sum_{\bm k'}W_{\bm k' \bm k}^0 = \sum_{\bm k'}{2\pi\over\hbar}\delta(\varepsilon_{k}-\varepsilon_{k'}) \: 2\Xi_c^2 {k_\text{B}T\over 2\rho s^2} = {2\pi\over\hbar}g \Xi_c^2 {k_\text{B}T\over \rho s^2},
\end{equation}
we have
\begin{equation}
W^{(as)}_{\bm k' \bm k} = -\xi {1\over \tau_{ph}} {\Xi_v \over \Xi_c}
 {\hbar^2 \over 2mg} \left[-\delta(\varepsilon_{k'}-\varepsilon_{k})\Phi_0+ \delta'(\varepsilon_{k'}-\varepsilon_{k})(\Phi_1 - \Phi_2) \right] F_{{\rm drag},x}
.
\end{equation}
We took here into account Eq.~\eqref{e:vect}.

Having calculated the asymmetric scattering probability we can now calculate the VHE from the kinetic equation. 
The anisotropic correction to the distribution function is found from
\begin{equation}
{\delta f_{\bm k}\over \tau_p} = \sum_{\bm k'}W^{(as)}_{\bm k \bm k'} f_0(\varepsilon_{k'}).
\end{equation}
The VHE current is given by
\begin{equation}
j_y = e \sum_{\bm k} \delta f_{\bm k} v_{\bm k,y} = e\tau_p\sum_{\bm k,\bm k'} v_{\bm k,y} W^{(as)}_{\bm k \bm k'} f_0(\varepsilon_{k'}) {\equiv} e\tau_p\sum_{\bm k,\bm k'} v_{\bm k',y} W^{(as)}_{\bm k' \bm k} f_0(\varepsilon_{k}).
\end{equation}
Therefore we need the averages $\left< \Phi_{0,1,2}v_{\bm k',y}\right>_{\varphi_{\bm p},\varphi_{\bm k},\varphi_{\bm k'}}$:
\begin{equation}
j_y =-eF_{{\rm drag},x}\xi {\tau_p\over \tau_{ph}} {\Xi_v \over \Xi_c}
 {\hbar^2\over 2m g}\sum_{\bm k,\bm k'} v_{\bm k',y} \left[-\delta(\varepsilon_{k'}-\varepsilon_{k})\Phi_0+ \delta'(\varepsilon_{k'}-\varepsilon_{k})(\Phi_1 - \Phi_2) \right] f_0(\varepsilon_{k}).
\end{equation}

Using the relations
\begin{align}
[\bm k' \times \bm p + \bm k \times \bm p' +  \bm p \times \bm k +  \bm p' \times \bm k']_z
= 2[\bm k' \times \bm p + \bm p \times \bm k +  \bm k \times \bm k']_z, \nonumber \\
\varepsilon_{\bm k+\bm k' - \bm p}-\varepsilon_k = {\hbar^2\over 2m}(k'^2+p^2+2\bm k\cdot \bm k' - 2\bm k\cdot \bm p-2\bm k'\cdot \bm p),
\end{align}
we obtain
\begin{equation}
\left< \Phi_{0}v_{\bm k',y}\right>_{\varphi_{\bm p},\varphi_{\bm k},\varphi_{\bm k'}}
= {m\over \hbar^3}\left<
{\sin{\varphi_1}[-\sin{\varphi_1}+\sin{\varphi_2} +  \sin{(\varphi_1-\varphi_2)}] \over [1+\cos{(\varphi_1-\varphi_2)} - \cos{\varphi_2}- \cos{\varphi_1}]^2}\right>_{\varphi_1,\varphi_2}.
\end{equation}
Here we introduced $\varphi_1=\varphi_{\bm k'}-\varphi_{\bm p}$, $\varphi_2=\varphi_{\bm k}-\varphi_{\bm p}$.
Similarly,
\begin{equation}
\left< \Phi_{1}v_{\bm k',y}\right>_{\varphi_{\bm p},\varphi_{\bm k},\varphi_{\bm k'}}
={1\over \hbar}k'^2x^2 \left<
{\sin{\varphi_1}[-\sin{\varphi_1}+x\sin{\varphi_2} +  \sin{(\varphi_1-\varphi_2)}] \over (1+x^2)/2+x\cos{(\varphi_1-\varphi_2)} - x^2\cos{\varphi_2}-x\cos{\varphi_1}}\right>_{\varphi_1,\varphi_2},
\end{equation}
where we introduced $x=k/k'$.
Finally,
\begin{equation}
\left< \Phi_{2}v_{\bm k',y}\right>_{\varphi_{\bm p},\varphi_{\bm k},\varphi_{\bm k'}}
={1\over \hbar}k'^2\left<
{[\sin{\varphi_1}- x\sin{(\varphi_1-\varphi_2)}] [-\sin{\varphi_1}+x\sin{\varphi_2} +  x\sin{(\varphi_1-\varphi_2)}] \over 1+x\cos{(\varphi_1-\varphi_2)} - x\cos{\varphi_2}-\cos{\varphi_1}}\right>_{\varphi_1,\varphi_2}.
\end{equation}
Therefore we have
\begin{equation}
j_y =-{1\over \hbar}eF_{{\rm drag},x}\xi {\tau_p\over \tau_{ph}} {\Xi_v \over \Xi_c}
 {\hbar^2\over 2m g}\sum_{\bm k,\bm k'} k'^2 [-\delta(\varepsilon_{k}-\varepsilon_{k'}) G_0/2\varepsilon_{k'} + \delta'(\varepsilon_{k}-\varepsilon_{k'}) G(k/k')]f_0(\varepsilon_{k}),
\end{equation}
where
\begin{equation}
G_0=\left<
{\sin{\varphi_1}[-\sin{\varphi_1}+\sin{\varphi_2} +  \sin{(\varphi_1-\varphi_2)}] \over [1+\cos{(\varphi_1-\varphi_2)} - \cos{\varphi_2}- \cos{\varphi_1}]^2}\right>_{\varphi_1,\varphi_2},
\end{equation}
\begin{multline}
G(x)=
 \biggl<
{x^2\sin{\varphi_1}[-\sin{\varphi_1}+x\sin{\varphi_2} +  \sin{(\varphi_1-\varphi_2)}] \over (1+x^2)/2+x\cos{(\varphi_1-\varphi_2)} - x^2\cos{\varphi_2}-x\cos{\varphi_1}} \\
-
{[\sin{\varphi_1}- x\sin{(\varphi_1-\varphi_2)}] [-\sin{\varphi_1}+x\sin{\varphi_2} +  x\sin{(\varphi_1-\varphi_2)}] \over 1+x\cos{(\varphi_1-\varphi_2)} - x\cos{\varphi_2}-\cos{\varphi_1}}
\biggr>_{\varphi_1,\varphi_2}.
\end{multline}

Summation over $\bm k, \bm k'$ is performed as follows:
\begin{multline}
\bm j =-{e\over \hbar} [\hat{\bm z}\times \bm F_{\rm drag}]\xi {\tau_p\over \tau_{ph}} {\Xi_v \over \Xi_c}
 {1\over g}\sum_{\bm k,\bm k'} [-\delta(\varepsilon_{k}-\varepsilon_{k'}) G_0/2 + \varepsilon_{k'}\delta'(\varepsilon_{k}-\varepsilon_{k'}) G(k/k')] f_0(\varepsilon_{k}) \\
=-{e\over \hbar} [\hat{\bm z}\times \bm F_{\rm drag}]\xi {\tau_p\over \tau_{ph}} {\Xi_v \over \Xi_c}
 N\left\{ G(1)-{1\over 2} [G_0+G'(1)]\right\}.
\end{multline}
Calculations show that $G(1)=1$ and $G_0+G'(1)=1$. This yields 
\begin{equation}
\bm j = -{1\over 2}{e\over \hbar}\xi  N {\tau_p\over \tau_{ph}} {\Xi_v \over \Xi_c}
[\hat{\bm z}\times \bm F_{\rm drag}].
\end{equation}

We see that the coherent contribution for electrons at phonon drag is a quarter of that in electric field. The same relation holds for excitons. Therefore we finally have
\begin{equation}
\label{i:vhe:skew:drag}
\chi^{(skew,d)} = \frac{\chi^{(skew,2ph,s)}}{2}+ \frac{\chi^{(skew,coh,s)}}{4}.
\end{equation}
It proves Eq. (20) of the main text.

{\section{Exciton VHE caused by their  density and  temperature gradients}\label{sec:diffusion:Nernst}}

{While in the main text we focused on the VHE caused either by a synthetic field or by the phonon drag effect, i.e., the gradient of the lattice temperature, it is instructive to consider also other relevant situations where the excitons are driven either by their density gradient (i.e., in the course of diffusion) or by their temperature gradient (i.e., due to the Seebeck or thermal drift effect), option (iii) discussed in the main text, page 2. The key theoretical challenge is the identification of the force acting on the excitons since in both cases, formally, there is no real potential gradient which makes the excitons moving. In this situation, however, it is possible to invoke general arguments~\cite{PhysRev.135.A1505} (see also Refs.~\cite{PhysRevLett.97.026603,PhysRevLett.114.196601,PhysRevB.101.155204,PhysRevB.101.195126} for discussion of limitations) and identify the exciton density and temperature gradients with some effective fields.}

{We start from the case of the density gradient, where the excitons are diffusing. In this case one can formally introduce a driving force 
\begin{equation}
\label{F:d:diff}
\bm F_d^{(diff)} = - \bm \nabla \mu,
\end{equation}
where $\mu\equiv \mu(\bm R)$ is the chemical potential of excitons and the associated (synthetic) ``electro-chemical'' potential energy 
\begin{equation}
\label{electro:chem}
U_{tot}(\bm R) = U(\bm R) + \mu(\bm R), 
\end{equation}
where $U(\bm R)$ is given by Eq. (6) of the main text. In this case the resulting exciton valley Hall current and the parameter $\chi$ in Eq. (1) of the main text can be immediately calculated from Eqs. (13), (16), and (17). In this case, there is absolutely no difference between the synthetic field and the chemical potential gradient. Particularly, the cancellation of the anomalous velocity contribution with a part of the side-jump  one is present.}

{The situation is more subtle with the temperature gradient where the valley Hall current is caused by the Seebeck (thermal drift) effect, i.e., where $T\equiv T(\bm R)$ and we set $\mu = const$. In this case, formally, one has to introduce an effective scalar potential or vector potential which couples to the energy or energy flux density, respectively~\cite{PhysRev.135.A1505,PhysRevLett.114.196601,PhysRevB.101.155204}. On a technical side, however, it is convenient to use the Onsager relations and start from the phenomenological form of the transport equations in a single valley [cf. Refs~\cite{PhysRev.135.A1505,ll8_eng}]:
\begin{subequations}
\label{phenom:gen}
\begin{align}
i_\alpha = a_{\alpha\beta} \frac{F_\beta}{T} + b_{\alpha\beta} \frac{\partial}{\partial x_\beta} \frac{1}{T}\label{flux:exc}\\
q_\alpha - U i_\alpha = c_{\alpha\beta} \frac{F_\beta}{T} + d_{\alpha\beta} \frac{\partial}{\partial x_\beta} \frac{1}{T}\label{flux:heat}.
\end{align}
\end{subequations}
Here $\bm i$ is the flux of excitons and $\bm q$ is the heat flux, $\alpha,\beta=x,y$ are the Cartesian subscripts and we set $k_B=1$ here for brevity. The material tensors $\hat a$ and $\hat b$ describe the current generation due to the force field and temperature gradient, while the tensors $\hat c$ and $\hat d$ describe the heat current generation as a response to the driving force and temperature gradient. Particularly, the components $b_{xy}$, $b_{yx}$ describe the valley Nernst  
effect (valley Hall current driven by the temperature gradient) in question. In Eqs.~\eqref{phenom:gen} we have included $1/T$ factors in the field terms and $\bm \nabla T^{-1}$ (instead of $\bm \nabla T$ for further convenience) and subtracted $U\bm i$ from the heat flux in order to exclude the energy $U$ transferred by each exciton~\cite{ll8_eng}. In accordance with the general principle of the symmetry of kinetic coefficients (Onsager relations) and taking into account the three-fold rotatational symmetry of the monolayer we obtain
\begin{subequations}
\label{Onsager}
\begin{align}
a_{xx} = a_{yy}, \quad a_{xy} = -a_{yx}, \quad d_{xx} = d_{yy}, \quad d_{xy} = -d_{yx}, \\
b_{xx} = b_{yy} = c_{xx} = c_{yy}, \quad b_{xy} = - c_{yx}, \quad b_{yx} = -c_{xy}.\label{Onsager:b}
\end{align} 
\end{subequations}
Equation~\eqref{Onsager:b} makes it possible, in particular, to evaluate the valley Nernst
effect via calculating the heat current 
\begin{equation}
\label{heat:curr}
\bm q = \sum_{\bm K} \bm v_{\bm K}(\mathcal E_{\bm K}-\mu) f_{\bm K},
\end{equation}
 as a response to a synthetic field $\bm F = -\bm \nabla U(\bm R)$  with $\bm v_{\bm K}$ being the exciton velocity, $\mathcal E_{\bm K}$ being its dispersion, and $f_{\bm K}$ being its non-equilibrium distribution function. As a result, one can immediately identify the anomalous velocity, side-jump, and skew scattering contributions to the components $c_{xy},c_{yx}$ and obtain the cancellation of the anomalous velocity with the part of the side-jump contribution. Automatically, by virtue of Eq.~\eqref{Onsager:b} same cancellations take place for the valley Nernst
 effect as well, i.e., for the components $b_{xy},b_{yx}$ of interest for us. As a result, the valley Hall current in the case of the Nernst
 effect is fully analogous to the current which arises in the presence of the synthetic force field.}
 
{In the relevant case of the non-degenerate excitons ($\mu<0$, $|\mu| \gg T$) and the short-range scattering with the energy-independent scattering time $\tau_p$  one can introduce a driving force
\begin{equation}
\label{F:Seebeck}
\bm F_d^{(Seebeck)} =  \frac{\mu}{T} \bm \nabla T,
\end{equation}
and explicitly check that the exciton valley Hall flux calculated via the effective potential energy associated with the force field, Eq.~\eqref{F:Seebeck} [after Eqs. (1), (13), (16), and (17) of the main text] corresponds to the rigorous result. For instance, using this procedure we obtain the contribution due to the anomalous velocity ($|\mu/T| \gg 1$)
\begin{equation}
\label{i:vhe:seebeck:anom}
\bm i_v^{({Nernst},anom)} = \frac{N}{2} \frac{\mu}{T} [\bm{\mathcal F}\times \bm {\nabla}  T] = \frac{N}{2} [\bm{\mathcal F}\times \bm F_d^{(Seebeck)}],
\end{equation}
in agreement with Refs.~\cite{PhysRevLett.97.026603,PhysRevB.101.155204}.
} 

{Thus, the exciton valley Hall effect under the density gradient and temperature gradient can be calculated using the expressions Eqs. (1), (13), (16), and (17) of the main text for a synthetic force field with appropriate force, Eq.~\eqref{F:d:diff} and \eqref{F:Seebeck}, respectively.}

\section{Drift-diffusion equations}

\subsection{Derivation}

Here we derive drift-diffusion equations describing valley separation of excitons due to the VHE. To that end, we follow Ref.~\cite{dyakonov_book} and SI of Ref.~\cite{PhysRevLett.120.207401} {and consider a slow, as compared with the exciton momentum relaxation time $\tau_p$, dynamics of the particles}. We introduce the following notations: Let $N_\pm$ be the occupancies (densities) of excitonic states active in $\sigma^+$ and $\sigma^-$ polarizations (we denote these excitons as excitons in $\bm K_\pm$ valleys according to the valley degree of freedom of electron in the exciton),
\begin{subequations}
\begin{equation}
\label{dens}
N = N_+ + N_-,
\end{equation}
is the total exciton density. The $z$-component of exciton pseudospin $S_z$ is given by
\begin{equation}
\label{Sz}
S_z = \frac{N_+ - N_-}{2}.
\end{equation}
\end{subequations}
Note that $N_\pm$, $N$ and $S_z$ depend on time $t$ and position $\bm r$. The partial flux densities of excitons in $\bm K_\pm$ valleys are denoted as $\bm i_\pm$, thus the total exciton flux density reads
\begin{subequations}
\begin{equation}
\label{flux}
\bm i = \bm i_+ + \bm i_-,
\end{equation}
and the exciton valley flux density is defined as
\begin{equation}
\label{VHE:flux}
\bm i_v = \frac{\bm i_+ - \bm i_-}{2},
\end{equation}
\end{subequations}
in accordance with the main text.

In the presence of the drag force $\bm F_d$ (either due to the synthetic field or due to the phonon drag) the exciton flux in a given valley reads {[cf. Eq.~\eqref{flux:exc}]}
\begin{equation}
\label{flux:drag}
\bm i_\pm = \frac{\tau_p}{M} \bm F_d N_\pm \pm 2 \chi {N_\pm} [\hat{\bm z} \times \bm F_d],
\end{equation}
with $\tau_p$ being the momentum relaxation time and appropriate  constant $\chi$ which depends, as shown above, on the origin of $\bm F_d$. {As before, we consider the linear transport regime where exciton fluxes are proportional to the first power of the drag force, $\bm F_d$. Note that under conditions of optical experiments the excitons can form a non-equilibrium distribution, in which case the parameters $\tau_p$ and $\chi$ can depend on this distribution (e.g., via the effective temperature or kinetic energy of excitons). In this case, an additional equation describing the energy balance in the system or distribution function of excitons is needed, i.e., the system of equations for $N_\pm$ (presented below) should be supplemented by Eq.~\eqref{flux:heat} for the heat current and an equation for the temperature. In what follows to simplify the analysis we assume, however, that a thermal equilibrium with an effective exciton temperature $T$ is reached.} Equations~\eqref{flux:drag} describe the fluxes induced by the force, i.e., the drift of the excitons. The spatial gradients of $N_+$ and $N_-$ produce diffusive contributions to the valley fluxes $\bm i_\pm^{(diff)} = - D \bm \nabla  N_\pm$ with $D$ being the diffusion coefficient of excitons. Accordingly, the fluxes in the presence of both the force field and the density gradients read
\begin{subequations}
\label{fluxes:full}
\begin{align}
\label{flux:exc:full}
\bm i = \frac{\tau_p}{M} \bm F_d N + 4\chi[\hat{\bm z} \times \bm F_d] S_z {-4\frac{MD}{\tau_p}\chi^{(s)}[\hat{\bm z} \times \bm \nabla]S_z} - D \bm \nabla N,\\
\bm i_v =\frac{\tau_p}{M} \bm F_d S_z+ {\chi} [\hat{\bm z} \times \bm F_d]N {-\frac{MD}{\tau_p} {\chi^{(s)}} [\hat{\bm z} \times \bm \nabla]N} - D\bm \nabla S_z.\label{flux:val:full}
\end{align}
\end{subequations}
{Note that here $\chi$ depends on the mechanism of the drag force, while $\chi^{(s)}$ is that for the diffusion mechanism which, as noted in Sec.~\ref{sec:diffusion:Nernst} is equivalent to action of synthetic field. The terms with $\chi^{(s)}$, however, do not contribute to the drift-diffusion equations, see below.}

Making use of the continuity equations and taking into account the finite lifetime of the particles we arrive at the set of the drift-diffusion equations [Eqs. (21) of the main text]
\begin{subequations}
\label{drift:diff}
\begin{align}
&\frac{\partial N}{\partial t} = D\Delta N - \frac{\tau_p}{M} \bm F_d \cdot \bm \nabla N - 4\chi   [\bm F_{d} \times \bm \nabla]_z S_z - \frac{N}{\tau_0},\\
&\frac{\partial S_z}{\partial t} = D\Delta S_{z} - \frac{\tau_p}{M} \bm F_d \cdot \bm \nabla S_z - \chi  [\bm F_{d} \times \bm \nabla N]_z -  \frac{S_z}{\tau_s}.
\end{align}
\end{subequations}
Note that these equations can be recast in somewhat more symmetric form if one, instead of $S_z$ introduces the valley polarization density $P_z = 2S_z$ and the parameter $\beta = 2\chi$ [cf. Ref.~\cite{dyakonov_book}]. In these notations Eqs.~\eqref{drift:diff} read
\begin{subequations}
\label{drift:diff:1}
\begin{align}
&\frac{\partial N}{\partial t} = D\Delta N - \frac{\tau_p}{M} \bm F_d \cdot \bm \nabla N - \beta  [\bm F_{d} \times \bm \nabla]_z P_z - \frac{N}{\tau_0}, \label{eq:Nr}\\
&\frac{\partial P_z}{\partial t} = D\Delta P_{z} - \frac{\tau_p}{M} \bm F_d \cdot \bm \nabla P_z -  \beta  [\bm F_{d} \times \bm \nabla N]_z -  \frac{P_z}{\tau_s}.\label{eq:Sr}
\end{align}
\end{subequations}

{\subsection{Limiting cases}}

{Here we consider two important limiting cases which admit simple solutions of the drift-diffusion equations~\eqref{drift:diff} or \eqref{drift:diff:1} and illustrate physics of the exciton valley separation.}

{Let us start from the model case where the excitons are solely driven by the \emph{diffusion}. In this situation only diffusive fluxes are present and $\bm F_d$ in Eqs.~\eqref{fluxes:full} and \eqref{drift:diff} should be set to zero:
\begin{subequations}
\label{diffusion:only}
\begin{align}
&\frac{\partial N}{\partial t} = D\Delta N - \frac{N}{\tau_0},\\
&\frac{\partial S_z}{\partial t} = D\Delta S_{z}-  \frac{S_z}{\tau_s}. \label{S:z:diff}
\end{align}
\end{subequations} 
Here, the valley polarization can arise only due to the inhomogeneity of the sample or in the vicinity of its edge where the drift valley Hall current should be compensated by the diffusion of the valley polarization, similarly to the spin accumulation at the sample edges in the case of the spin-Hall effect~\cite{dyakonov_book}. The same result holds in the case where the excitons are driven by their temperature gradient with $\bm\nabla T \parallel \bm \nabla N$. One can also claim that the absence of valley polarization in the bulk of the sample despite the gradients of temperature and density is a general consequence of the symmetry: There is no way to build a pseudovector component $S_z$ from the scalar $N$ and vector $\bm \nabla N$ in the systems with $D_{3h}$ point symmetry~see SI to Ref.~\cite{PhysRevLett.120.207401}. That is why, in Ref.~\cite{Onga:2017aa} the effect is most probably related to the inhomogeneity of the sample and its finite size.}

{Next, we analyze the situation where the excitation is homogeneous and overall temperature gradient or other drag force is present in the whole sample. We demonstrate that in this situation the valley polarization appears at the sample edges oriented across $\bm F_d$. To be specific let $\bm F_d \parallel x$ (the specific origin of $\bm F_d$ affects on the the parameter $\chi$ or $\beta$ in the sets~\eqref{drift:diff} or \eqref{drift:diff:1} of drift-diffusion equations, see the main text and discussion above). Let the sample be infinite along the $x$-axis to avoid discussing the accumulation of excitons themselves and compensation of their drift current, but consider the sample bounded at $y=\pm L/2$, with $L$ being the sample width. In this geometry Eq.~\eqref{S:z:diff} holds, but it should be supplemented by the boundary conditions which require the valley flux to vanish at $y=\pm L/2$, thus, in accordance with Eq.~\eqref{flux:val:full}:
\begin{equation}
\label{bc}
\chi N F_d - D \frac{\partial S_z}{\partial y} =0 \quad \mbox{at} \quad y=\pm L/2.
\end{equation} 
Assuming that $L \gg \sqrt{D\tau_s}$  we solve the set of Eqs.~\eqref{S:z:diff} and \eqref{bc} with the result
\begin{equation}
S_z(y{\approx \pm L/2}) \approx \pm \frac{\chi N F_d L_s}{D} \exp{\left(-\frac{|y {\mp} L/2|}{L_s}\right)}, \quad L_s = \sqrt{D\tau_s}.
\end{equation}
This situation is quite analogous to the spin accumulation at the sample edges under the conditions of the electron spin Hall effect~\cite{dyakonov_book}.}

{Experimental techniques for optical generation of excitons and detection of their propagations allow one to consider the situation where the excitons are created locally and their propagation in space is monitored in the real time, while the synthetic force field or temperature gradient is created homogeneously in the sample. In this case, an interesting situation arises where the valley polarization takes place within the cloud of propagating excitons. This situation is analyzed in the main text and the analytical solution and further details are summarized below.}

\subsection{Analytical solution}

The set of Eqs.~\eqref{drift:diff} admits analytical solution. We consider the situation where unpolarized excitons were created at $t=0$, let $N_0(\bm r)$ be the initial density profile. Performing the Fourier-transform of Eqs.~\eqref{drift:diff} and taking into account the initial condition we arrive at the set of the algebraic equations
\begin{subequations}
\label{drift:diff:F}
\begin{align}
&\left[-\mathrm i\omega + D q^2+ \mathrm i \frac{\tau_p}{M} (\bm F_d\cdot \bm q)+\tau_0^{-1}\right]N_{\omega,\bm q} +  4\mathrm i \chi   [\bm F_{d} \times \bm q]_z S_{z,\omega,\bm q} = N_{0,\bm q} \label{eq:N},\\
&\left[-\mathrm i\omega + D q^2+ \mathrm i \frac{\tau_p}{M} (\bm F_d\cdot \bm q)+\tau_s^{-1}\right]S_{z,\omega,\bm q} = -\mathrm i \chi  [\bm F_{d} \times \bm q ]_zN_{\omega,\bm q}.\label{eq:S}
\end{align}
\end{subequations}
Here $N_{0,\bm q}$ is the Fourier-transform of the $N_0(\bm r)$ over the coordinate, $N_{\omega,\bm q}$ and $S_{z,\omega,\bm q}$ denote the Fourier transforms of $N(\bm r,t)$ and $S_z(\bm r,t)$ over time and coordinate.

\begin{figure}[h]
\includegraphics[width=0.8\linewidth]{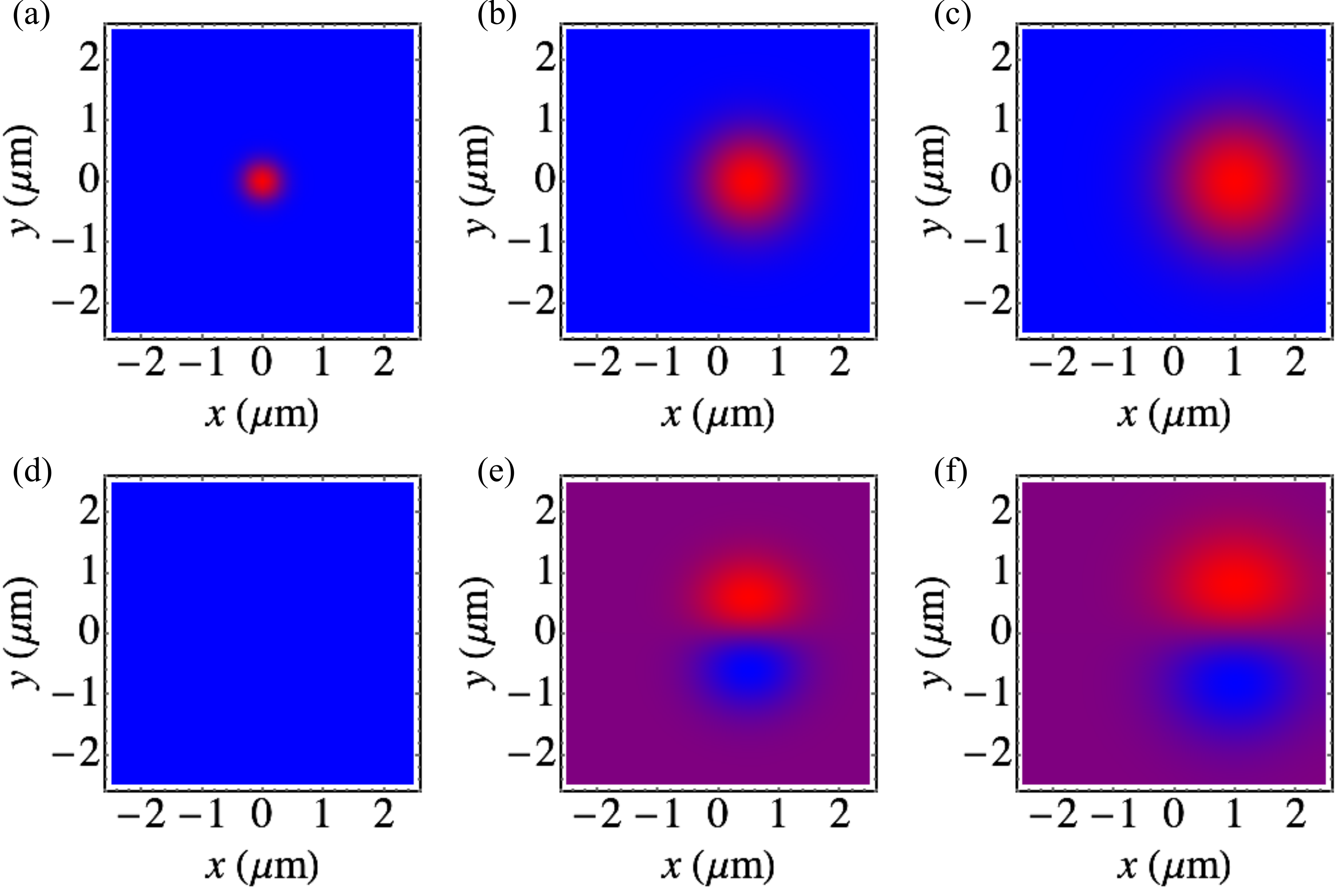}
\caption{Exciton density profile at $t=0$ [panel (a)], $t=0.5$~ns [panel (b)], and $t=1$~ns [panel (c)] calculated after Eqs.~\eqref{drift:diff} in the presence of $\bm F_{d} \parallel x$. Panels (d-f) show the exciton pseudospin $S_z$. Parameters: $D=3$~cm$^2/$s, $\tau_0=1$~ns, $\tau_s=0.3$~ns, drift velocity $F_d \tau_p/M=1$~$\mu$m/ns, valley Hall angle $\beta=2\chi M/\tau_p=0.1$, initial spot radius $r_0=0.33$~$\mu$m. False color scale is used to highlight the density and pseudospin spatial profiles.}\label{fig:6panel}
\end{figure}

\begin{figure}[h]
\includegraphics[width=0.8\linewidth]{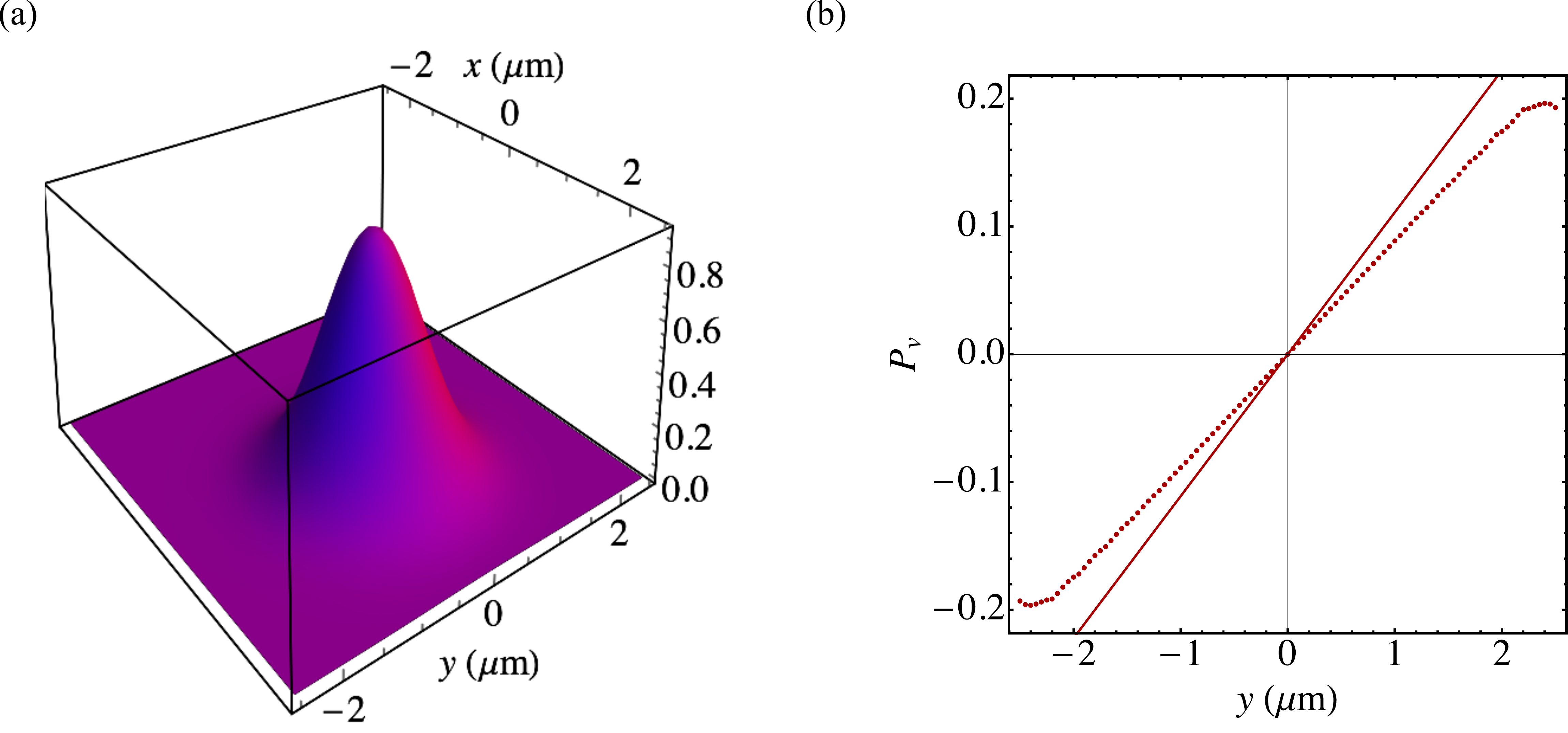}
\caption{(a) Exciton density profile at $t=0.5$~ns calculated after Eqs.~\eqref{drift:diff} in the presence of $\bm F_{d} \parallel x$. Color shows exciton pseudospin $z$-component, $S_z$. (b) Exciton polarization degree $P_v$ calculated numerically (points) and using analytical asymptotics, Eq.~\eqref{Pv}. Parameters are the same as in Fig.~\ref{fig:6panel}. }\label{fig:profile}
\end{figure}

Under reasonable assumptions $\chi N/\tau_p \ll 1$, $\tau_s \ll \tau_0$ and for typical times $t \gtrsim \tau_s$ one can disregard the term $\propto S_z$ in Eq.~\eqref{eq:N}, while in Eq.~\eqref{eq:S} keep, in the left hand side, only $\tau_s^{-1}$. As a result
\[
S_{z,\omega,\bm q} =  -\mathrm i \tau_s \chi  [\bm F_{d} \times \bm q ]_zN_{\omega,\bm q},
\]
or transforming to the real space 
\begin{equation}
\label{Sz:rt}
S_z(\bm r, t) = -\chi \tau_s [\bm F_d \times \bm \nabla N(\bm r,t)]_z .
\end{equation}
Note that this expression can be directly derived from Eq.~\eqref{eq:Sr} neglecting time and spatial derivatives of $S_z$. The solution of Eq.~\eqref{eq:N} with $\chi S_z$ term neglected can be readily written in the closed analytical form. For example, we consider the initial Gaussian distribution of excitons
\begin{equation}
\label{Gauss}
N_0(\bm r) = \frac{C}{\pi r_0^2} \exp{(-r^2/r_0^2)},
\end{equation}
with the constant $C$ and $r_0$ being the initial spot radius. Introducing the drift velocity $\bm v_d = \bm F_d \tau_p/M$ we arrive at
\begin{equation}
\label{Nrt}
N(\bm r, t) = \sum_{\bm q} \exp{\left[-(Dq^2+\tau_0^{-1}) t +\mathrm i \bm q (\bm r -  \bm v_d t)\right]} N_{0,\bm q} = \frac{C}{\pi r_0^2 + 4 D t}\exp{\left(-\frac{(\bm r-\bm v_d t)^2}{r_0^2+4D t}-\frac{t}{\tau_0}\right)}.
\end{equation}
The valley polarization degree ($t \gg r_0^2/D$)
\begin{equation}
\label{Pv}
P_v = \frac{2S_z(\bm r,t)}{N(\bm r,t)} \approx  \frac{\chi \tau_s}{Dt} [\bm F_d \times \bm r]_z, 
\end{equation}
in agreement with numerical calculations presented in the main text and in Figs.~\ref{fig:6panel} and \ref{fig:profile}.

\newpage